\documentclass[aps,prd,twocolumn,superscriptaddress,nofootinbib]{revtex4}
\usepackage{hyperref}
\usepackage{amsmath,amssymb,amsfonts}
\usepackage{color}
\usepackage{graphicx}
\usepackage{enumerate} 
\usepackage{colordvi} 
\usepackage{bm}
\usepackage{here}

\newcommand{\mnras}{Mon. Not. Roy. Astron. Soc. }
\newcommand{\apjl}{Astrophys. J. Lett. }

\newcommand{\jcap}{J. Cosmol. Astropart. Phys. }
\newcommand{\physrep}{Phys. Rept. }
\newcommand{\aap}{Astron. Astrophys. }

\newcommand{\ssr}{Space Sci. Rev.}
\newcommand{\pasj}{Publ. Astron. Soc. Jpn.}

\newcommand{\vC}{\mathbf{C}}
\newcommand{\vq}{\mathbf{q}}
\newcommand{\vp}{\mathbf{p}}

\newcommand{\vx}{\mathbf{x}}

\newcommand{\vk}{\mathbf{k}}

\newcommand{\vecr}{\mathbf{r}}

\newcommand{\vv}{\mathbf{v}}

\newcommand{\wt}{\widetilde}

\newcommand{\himpc}{{\hbox {$~h^{-1}$}{\rm ~Mpc}}}

\newcommand{\higpc}{{\hbox {$~h^{-1}$}{\rm ~Gpc}}}

\newcommand{\hsimpcs}{{\hbox {$~h^{-2}$}{\rm ~Mpc}^2}}

\newcommand{\himsun}{{\hbox {$~h^{-1}$}{M_\odot}}}

\newcommand{\be}{\begin{equation}}
\newcommand{\ee}{\end{equation}}
\newcommand{\bey}{\begin{eqnarray}}
\newcommand{\eey}{\end{eqnarray}}
\newcommand{\bea}{\begin{array}}
\newcommand{\eea}{\end{array}}

\newcommand{\nn}  {\nonumber}
\newcommand{\regpt}{R{\footnotesize EG}PT }
\newcommand{\camb}{\texttt{\footnotesize CAMB} }
\newcommand{\rockstar}{\texttt{\footnotesize ROCKSTAR} }
\newcommand{\darkquest}{\texttt{D{\footnotesize ARK} Q{\footnotesize UEST} }}

\begin{document}
\title{
Intrinsic alignment statistics of density and velocity fields at large scales: Formulation, modeling, and 
baryon acoustic oscillation features 
}
\author{Teppei Okumura}\email{tokumura@asiaa.sinica.edu.tw}
\affiliation{Institute of Astronomy and Astrophysics, Academia Sinica, 11F of AS/NTU Astronomy-Mathematics Building, No. 1, Section 4, Roosevelt Road, Taipei 10617, Taiwan}
\affiliation{Kavli Institute for the Physics and Mathematics of the Universe (WPI), UTIAS, The University of Tokyo, Kashiwa, Chiba 277-8583, Japan}

\author{Atsushi Taruya}
\affiliation{Center for Gravitational Physics, Yukawa Institute for Theoretical Physics, Kyoto University, Kyoto 606-8502, Japan}
\affiliation{Kavli Institute for the Physics and Mathematics of the Universe (WPI), UTIAS, The University of Tokyo, Kashiwa, Chiba 277-8583, Japan}

\author{Takahiro Nishimichi}
\affiliation{Center for Gravitational Physics, Yukawa Institute for Theoretical Physics, Kyoto University, Kyoto 606-8502, Japan}
\affiliation{Kavli Institute for the Physics and Mathematics of the Universe (WPI), UTIAS, The University of Tokyo, Kashiwa, Chiba 277-8583, Japan}

\date{\today}

\begin{abstract}
The kinematic Sunyaev-Zel'dovich effect enables us to directly probe the density-weighted velocity field up to very large cosmic scales. We investigate the effects of intrinsic alignments (IAs) of dark-matter halo shapes on cosmic density and velocity fields on such large scales. In the literature IAs have been detected up to $\sim 100\himpc$ using the gravitational shear-intrinsic ellipticity correlation and the alignment correlation function. In this paper we introduce the corresponding various velocity statistics: the (density-weighted) velocity-intrinsic ellipticity correlation as well as the alignment pairwise infall momentum, momentum correlation function, and density-weighted pairwise velocity dispersion. We derive theoretical expressions for these velocity alignment statistics for the first time based on the assumption that the density fluctuation is a Gaussian random field. Using large-volume, high-resolution $N$-body simulations, we measure the alignment statistics of density and velocity fields. The behaviors of IAs in the velocity statistics are similar to those in the density statistics, except that the halo orientations are aligned with the velocity field up to scales larger than those with the density field, $r\gg 100\himpc$, because of a factor of the wave number in the linear relation between the density and velocity fields in Fourier space, $v\propto \delta/k$. We show that the detected IAs of the velocity field can be well predicted by the linear alignment model. We also demonstrate that the baryon acoustic oscillation features can be detected in both the conventional and alignment velocity statistics. Our results indicate that observations of IAs with the velocity field on large scales can provide additional information on cosmological models, complementary to those with the density field.  
\end{abstract}
\maketitle

\flushbottom
\section{Introduction}
Intrinsic alignments (IAs), correlations of galaxy orientations/shapes with surrounding fields such as the mass overdensity, arise due to physical processes during the formation and evolution of galaxies. Thus, investigating IAs in principle enables one to probe galaxy formation and evolution. IAs have been extensively studied also in the context of potential systematic effects in weak-lensing surveys \cite{Heavens:2000,Croft:2000,Lee:2000}, and their contamination to the weak-lensing shear correlations has been investigated both theoretically \cite{Catelan:2001,Crittenden:2001,Hirata:2004,Smith:2005,Bridle:2007,Okumura:2009a,Zhang:2010,Schneider:2010,Blazek:2011,Blazek:2015,Meng:2018,Codis:2018,Yao:2019} and observationally \cite{Brown:2002,Heymans:2004,Mandelbaum:2006,Hirata:2007,Okumura:2009,Joachimi:2011,Li:2013,Singh:2015,Huang:2016,Tonegawa:2018,Samuroff:2019} (see, e.g., Refs. \cite{Schafer:2009,Troxel:2015,Joachimi:2015,Kirk:2015,Kiessling:2015} for reviews). 

Recently, some studies have been made focusing on IAs not as a contamination in the weak lensing surveys but as an additional cosmological probe to constrain primordial non-Gaussianity and measure baryon acoustic oscillations (BAOs) \cite{Chisari:2013}. Moreover, the detectability of the cross correlation between galaxy shapes and cosmic microwave background $B$-mode polarization induced by primordial gravitational waves has been discussed \cite{Schmidt:2012,Chisari:2014}. There was a further discussion of the imprint of inflation on galaxy IA \cite{Schmidt:2015,Chisari:2016,Kogai:2018,Yu:2019}. 

It is well known that the effect of IA depends strongly on the mass of host dark-matter halos \cite{Jing:2002,Xia:2017,Piras:2018}. Clusters of galaxies are thus ideal objects to address a fundamental question of up to what scales luminous objects are aligned with the matter distribution in the large-scale structure of the universe. While orientations of galaxies are known to be misaligned with those of the host halos \cite{van-den-Bosch:2002,Okumura:2009}, such a misalignment between shapes of clusters and their host halos is much smaller \cite{Despali:2017,Umetsu:2018,Okabe:2018}. Using cluster and brightest cluster galaxy samples, strong alignment signals have been detected up to scales $\sim 100\himpc$ in observation \cite{Hirata:2007,Okumura:2009,Li:2013,van-Uitert:2017}. 

So far, most of the studies on IAs has focused on the alignments of the major axes of galaxies relative to the overdensity field, such as the gravitational shear-intrinsic ellipticity (GI) correlation \cite{Hirata:2004} and alignment density correlation function \cite{Paz:2008,Faltenbacher:2009}. However, it is of fundamental importance to consider the IA relative to the cosmic velocity field for various reasons. Now the velocity field at cosmic scales can be measured through several ways, such as peculiar velocity \cite{Strauss:1995} and kinematic Sunyaev-Zel'dovich (kSZ) \cite{Sunyaev:1980,Hand:2012} surveys. Since the velocity field in Fourier space is expressed as $\vv(\vk) \propto (i\vk/k^2) \delta(\vk)$ in linear theory, one expects that the velocity correlation signal is amplified compared to the density counterpart on large scales due to the prefactor of $\vk/k^2$. This trend has been seen in the measured pairwise velocity power spectrum in kSZ surveys \cite{De-Bernardis:2017,Sugiyama:2018}. References. \cite{Schmidt:2010,Lam:2011} have studied how the large-scale velocity field is affected by the presence of primordial non-Gaussianity. A method to improve the local non-Gaussianity constraints has been proposed with the velocity information from the kSZ tomography \cite{Munchmeyer:2018}. In addition, it has been demonstrated that velocities of infalling objects around clusters can be potentially used to constrain modified gravity models \cite{Lam:2012,Zu:2013,Adhikari:2018}.

In the short article \cite{Okumura:2017a}, we have simultaneously analyzed the large-scale IA $>100\himpc$ in real and redshift space and boundaries of massive dark-matter halos at $\sim 1\himpc$ using the phase-space information. The article is unpublished and only the section on the splashback radius has been significantly extended and published in Ref.~\cite{Okumura:2018a}. In this paper, we extend the section on the large-scale IA in Ref.~\cite{Okumura:2017a} and present the detailed study of IA effects with the density and velocity fields. We introduce the velocity-intrinsic ellipticity (VI) correlation function as a natural extension of the GI correlation to phase space. We then define the alignment velocity correlation statistics, namely the alignment density-momentum and momentum-momentum correlation functions, as well as the pairwise velocity statistics, the alignment pairwise mean infall momentum and density-weighted pairwise velocity dispersion. We derive comprehensive expressions for these statistics in the linear regime by averaging the joint probability distribution of density, velocity and ellipticity fields. We obtain their explicit forms by utilizing the linear tidal alignment (LA) model where the intrinsic ellipticity of a galaxy is a linear function of the tidal field. The derived alignment velocity statistics are tested by comparing with the measurements from $N$-body simulations over a broad range of scales, from the quasinonlinear regime to scales beyond $100\himpc$.  

This paper is organized as follows. In Sec.~\ref{sec:theory}, we briefly review the statistics for IA with the density field, and then extend to those for IA with the velocity field. We derive the theoretical predictions for the density and velocity IA statistics in Sec. ~\ref{sec:theoretical_prediction}. Section \ref{sec:analysis} describes the $N$-body simulations as well as the measurements of the IA statistics. In Sec.~\ref{sec:comparison} we compare the derived theoretical predictions to the $N$-body measurements. We demonstrate that features of BAOs can indeed be seen in the velocity IA statistics in Sec.~\ref{sec:bao}. Our conclusions are given in Sec.~\ref{sec:conclusion}. The Appendix \ref{sec:formula} provides useful analytic formulas for deriving the theoretical predictions of the IA statistics for the Gaussian random fields.

Throughout the paper, the following cosmological parameters are assumed \cite{Planck-Collaboration:2016}: $\Omega_m=1-\Omega_\Lambda = 0.315$, $\Omega_b=0.0492$, $h=0.673$, $n_s=0.965$, and $\sigma_8=0.8309$. 


\section{Intrinsic alignment statistics}\label{sec:theory}

\begin{figure}[b]
\includegraphics[width=0.45\textwidth,angle=0,clip]{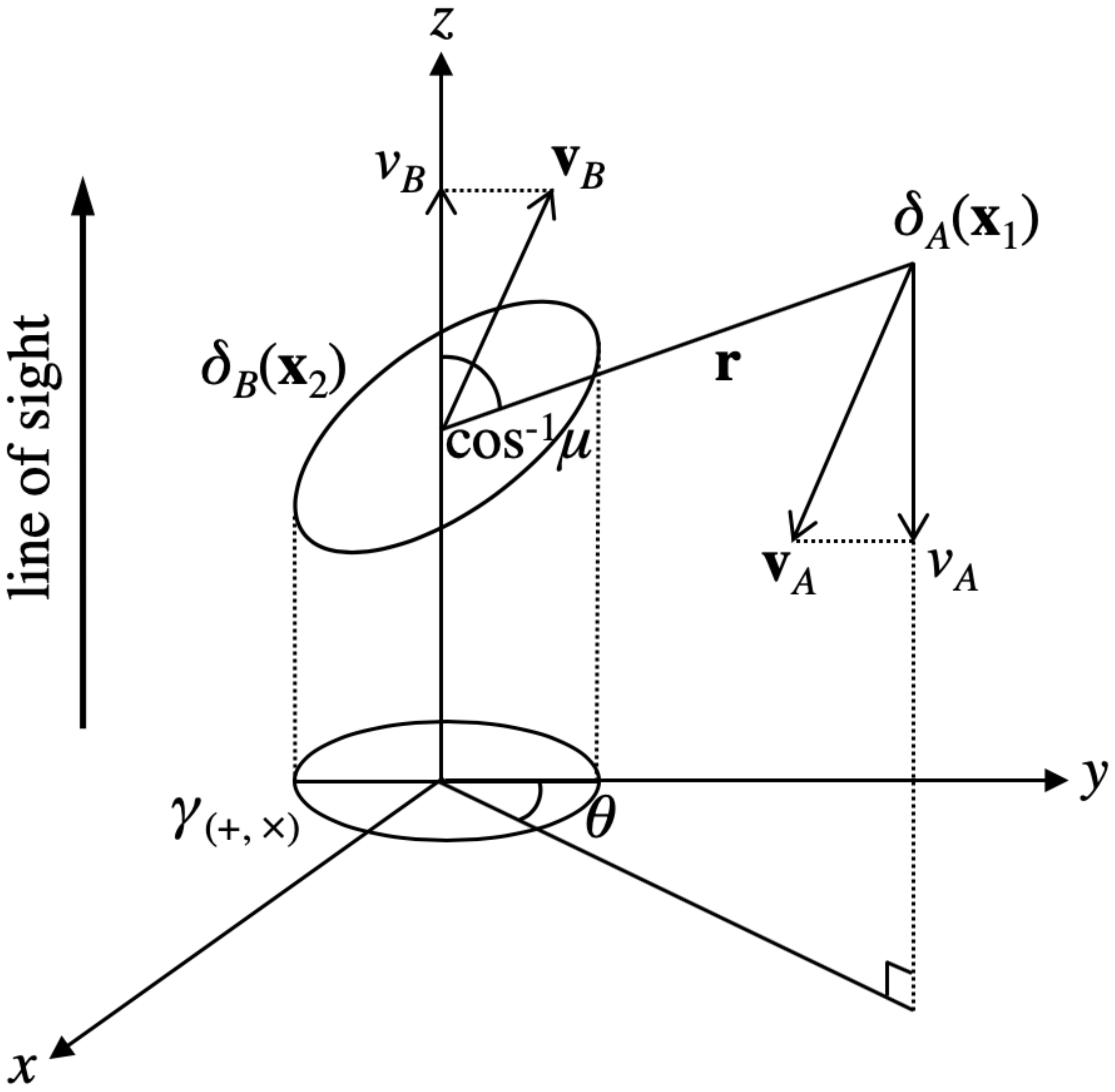}
\caption{Illustration of quantities used in this paper. The $z$ axis is the observer's line of sight and the $x$-$y$ plane corresponds to the celestial sphere.
}
\label{fig:ia}
\end{figure}

In this section we introduce the statistics used in this paper to quantify the IAs. They were presented in Refs. \cite{Okumura:2017a,Okumura:2018a}, but here we provide a more detailed derivation and explanation of the derived expressions. Throughout this paper, we consider the IA statistics in real space, taking the distant-observer or plane-parallel limit. An extension of including redshift-space distortions is rather straightforward, and will be reported in future work. Our notations for the quantities used to define the statistics are illustrated in Fig.~\ref{fig:ia}. We are particularly interested in the statistical correlation of the quantities associated with a pair of objects $A$ and $B$, taking special care with their relative orientation. In Fig.~\ref{fig:ia}, the objects $B$ are supposed to be halo or galaxy/cluster, and we assume that their shapes are measured on the celestial sphere. Based on this figure, we define various alignment statistics, which are all summarized in Table \ref{tab:statistics}. 

\begin{table*}[bt!]
\caption{Summary of the alignment statistics considered in this paper. The notation of each statistic is given by the column of $X(\vecr,\theta)$. The column of $\Xi(\vx_1,\vx_2,\theta)$ shows the part of $X$, expressed as $X(\vecr,\theta) = \left\langle \left[1+\delta_A ({\bf x}_1,\theta)\right]\left[1+\delta_B ({\bf x}_2)\right] \Xi(\vx_1,\vx_2) |\theta \right\rangle$ [see Eq.~(\ref{eq:operator})]. Note that as for the GI and VI correlations the $\theta$ dependence is averaged over by definition.}
\begin{center}
\begin{tabular}{lcccccc}
\hline\hline
 &  &  & Definition &  Formula & \multicolumn{2}{c}{Result (Fig.)}  \\ 
Statistics & $X(\vecr,\theta) $ & $\Xi(\vx_1,\vx_2)$ &   (Eq.) & (Eq.) & DM & Biased \\ 
\noalign{\hrule height 1pt}
GI correlation  & $1+\xi_{\delta_A+}(\vecr)$ & $\gamma_+(\vx_2)$ & (\ref{eq:gi}) & (\ref{eq:gi_la}) & \ref{fig:gi_m_r_2gpc_z014} & \ref{fig:gi_h_r_2gpc_z014} \\
VI correlation  & $\xi_{p_A+}(\vecr)$ & $v_A(\vx_1)\gamma_+(\vx_2)$ &   (\ref{eq:vi}) & (\ref{eq:vi_la}) & \ref{fig:gi_m_r_2gpc_z014} & \ref{fig:gi_h_r_2gpc_z014}\\[0.15cm] 
Density correlation & $1+\xi_{\delta_A\delta_B}(\vecr,\theta) $ & 1 & (\ref{eq:acf_density}) & (\ref{eq:acf_density_formula}) & \ref{fig:xidd_m} & \ref{fig:xidd_h}\\
Density-momentum correlation & $\xi_{\delta_Ap_B}(\vecr,\theta) $ & $v_B(\vx_2)$ & (\ref{eq:density_momentum}) & (\ref{eq:acf_infall_formula}) & &\\
Momentum-density correlation & $\xi_{p_A\delta_B}(\vecr,\theta) $ & $v_A(\vx_1,\theta)$ & (\ref{eq:momentum_density}) & (\ref{eq:acf_infall_formula}) &  &\\
Momentum correlation & $\xi_{p_Ap_B}(\vecr,\theta) $, $\psi_{AB}(\vecr,\theta) $ & $v_A(\vx_1,\theta)v_B(\vx_2)$ & (\ref{eq:momentum_momentum}) & (\ref{eq:acf_momentum_formula}) & \ref{fig:xidt_tt_vv_m} & \ref{fig:xidt_tt_vv_h} \\[0.15cm]
Pairwise infall momentum & $p_{AB}(\vecr,\theta) $ & $v_B(\vx_2)-v_A(\vx_1,\theta)$ & (\ref{eq:infall_momentum}) & (\ref{eq:acf_infall_formula}) & \ref{fig:xidt_tt_vv_m} &\ref{fig:xidt_tt_vv_h}\\
Density-weighted pairwise velocity dispersion & $\Sigma_{AB}^2(\vecr,\theta) $ & $(v_B(\vx_2)-v_A(\vx_1,\theta))^2$ & (\ref{eq:velocity_disp}) & (\ref{eq:velocity_disp_formula}) &\ref{fig:xidt_tt_vv_m} & \ref{fig:xidt_tt_vv_h}\\
\hline\hline
\end{tabular}
\end{center}
\label{tab:statistics}
\end{table*}

Let us define the intrinsic ellipticity, which is a key observable in the present paper. As shown in Fig.~\ref{fig:ia}, this quantity is defined on the celestial sphere, and is given as a two-component quantity, through the shape measurement of object $B$:
\bey
\gamma_{(+,\times)}^I(\vx)&=&\frac{1-(b/a)^2}{1+(b/a)^2}(\cos(2\theta),\sin(2\theta)) \nn \\
&\equiv&\gamma_0(\cos(2\theta),\sin(2\theta)). \label{eq:gamma_pc}
\eey
Here, $\theta$ is defined on the plane normal to the line-of-sight direction, and represents the angle between the projected separation vector pointing to the object $A$ from $B$ and the major axis of the shape of the object $B$ projected onto the celestial sphere. The ratio $b/a$ is the minor-to-major-axis ratio for the projected shape. 
The shape (shear) of a galaxy observed in a weak-lensing survey is always the sum of the lensing signal $\gamma^G$ and the intrinsic shape $\gamma^I$, namely $\gamma_{(+,\times)}=\gamma_{(+,\times)}^G+\gamma_{(+,\times)}^I$. Since we consider only the intrinsic shape and do not use the lensing components, we hereafter omit the superscript $I$ and simply write the intrinsic components as $\gamma_{(+,\times)}$.

\subsection{Density correlation}\label{sec:theory_density}
\subsubsection{Gravitational shear-intrinsic ellipticity correlation}

One of the commonly used statistics to describe IA is the cross correlation between the gravitational shear and intrinsic ellipticity (GI) \cite{Hirata:2004}, quantified by the density-ellipticity correlation. 
Given the two fields associated with objects $A$ and $B$ \footnote{Here we consider a general case where the two fields are different, e.g., $A$ and $B$ are the fields traced by galaxies and galaxy clusters, respectively, but the expression for the autocorrelation can be obtained by setting $A=B$.}, the GI correlation, $\xi_{\delta_A+}$, is defined by 
\be
1+\xi_{\delta_A+}(\vecr)=\left\langle[1+\delta_A(\vx_1)][1+\delta_B(\vx_2)]\gamma_+(\vx_2)\right\rangle, \label{eq:gi}
\ee
where $\vecr={\bf x}_2-{\bf x}_1$, and $\gamma_+$ is defined in the same way as shown in Fig.~\ref{fig:ia} through Eq.~(\ref{eq:gamma_pc}). The quantity $\delta_A$ is the perturbation of mass and number density fields, $\rho_A$, if the field $A$ is dark matter and biased objects (e.g., galaxies or clusters), respectively, where $\rho_A(\vx)=\bar{\rho}_A[1+\delta_A(\vx)]$ and $\bar{\rho}_A=\left\langle \rho_A({\bf x})\right\rangle$ is its mean value. Having a prefactor $(1+\delta_X)$ such as $\wt{\gamma}_+(\vx)\equiv [1+\delta_B(\vx)]\gamma_+(\vx)$ in Eq.~(\ref{eq:gi}) stands for a density-weighted quantity. The cross component, $\xi_{\delta_A\times}$, defined by replacing $\gamma_+(\vx_2)$ with $\gamma_\times(\vx_2)$ in Eq.~(\ref{eq:gi}), should be zero on all scales by symmetry.

\subsubsection{Alignment density-density correlation}

Here, we consider an alternative statistic to the GI correlation function, namely the alignment correlation function \cite{Paz:2008,Faltenbacher:2009,Papai:2013,Okumura:2017a,Osato:2018,Okumura:2018a}. This is defined as an extension of the conventional correlation function. 
The conventional two-point correlation function of the fields $A$ and $B$ is given by 
\be
\left\langle \rho_A({\bf x}_1)\rho_B({\bf x}_2)\right\rangle= \bar{\rho}_A\bar{\rho}_B [1+\xi_{\delta_A\delta_B}(r)]. \label{eq:2pcf}
\ee
Note that we are considering the real space where the statistical isotropy holds. Thus the function $\xi_{\delta_A\delta_B}$ depends only on the separation $r=|\vecr|$. 

To properly describe the alignment correlation function, let us first define the conditional average based on Fig.~\ref{fig:ia}. Consider the quantities $X$ and $Y$ made of the observables associated with objects $A$ and $B$, respectively. That is, the quantities $X$ and $Y$ are respectively defined at the positions $\vx_1$ and $\vx_2$, and we assume that the projected shape of the object $B$ has already been measured. Then, the conditional average of the quantities $X$ and $Y$ is defined to be the ensemble average for a fixed orientation angle $\theta$, as shown in Fig.~\ref{fig:ia}. For notational simplicity, dropping the arguments, $\vx_1$ and $\vx_2$, we write it as
\be
\langle X, \, Y | \theta\rangle \equiv \langle X(\vx_1) Y(\vx_2) | \theta \rangle ({\bf r},\theta). \label{eq:operator}
\ee
Note that inside the angle bracket on the left-hand side of Eq. (\ref{eq:operator}),  the quantity on the left (right) always implies the one associated with objects $A$ ($B$), and hence it is measured at $\vx_1$ ($\vx_2$).  By construction, the conditional average given in Eq.~(\ref{eq:operator}) has an explicit dependence on the angle $\theta$. Further it is in general given as a function of the separation vector ${\bf r}$, not $|{\bf r}|$. This is because the measured orientation angle of the halo or galaxy/cluster shape is defined on the plane normal to the line of sight, and this partly breaks statistical isotropy. We will see it explicitly in Sec.~\ref{sec:theoretical_prediction}. 

Provided the orientation-dependent average, we now define the alignment correlation function by replacing the standard ensemble average given in Eq.~(\ref{eq:2pcf}) with the one in Eq.~(\ref{eq:operator}). We have
\begin{align}
\left\langle \rho_A, \rho_B | \theta\right\rangle &= 
\bar{\rho}_A \bar{\rho}_B \left\langle 1+ \delta_A, 1 +\delta_B| \theta \right\rangle 
\nonumber
\\
&=\nonumber
\bar{\rho}_A \bar{\rho}_B \Bigl[ \langle 1, 1| \theta \rangle + \left\langle \delta_A, 1 | \theta \right\rangle + \left\langle 1,\delta_B | \theta \right\rangle  \\
&\nn \ \ \ \ \ \ \ \ \ \ \ \ \ \ \ \ \ \ \ \ \ \ \ \ \ \ \ \ \ \ \ \ \ \ \ \ +\left\langle \delta_A, \delta_B | \theta \right\rangle \Bigr] 
\\
&=\bar{\rho}_A \bar{\rho}_B \Bigl[1+ \left\langle \delta_A, 1 | \theta \right\rangle +\left\langle \delta_A, \delta_B | \theta \right\rangle \Bigr]. 
\label{eq:2pcf_theta}
\end{align}
Here we used the fact that $\langle 1, \delta_B|\theta\rangle = \langle\delta_B\rangle = 0$ because this is reduced to the standard ensemble average. In what follows, the same procedure will be applied to the calculations of IA statistics. 

Comparing Eq.~(\ref{eq:2pcf_theta}) to Eq.~(\ref{eq:2pcf}), we obtain the alignment density-density correlation, $\xi_{\delta_A\delta_B}(\vecr,\theta)$: 
\bey 
\xi_{\delta_A\delta_B}(\vecr,\theta)&=&\left\langle [1+\delta_A({\bf x}_1,\theta)][1+\delta_B({\bf x}_2)]\right\rangle -1 \nn \\
&=& \langle\delta_A,\delta_B|\theta\rangle + \langle\delta_A,1|\theta\rangle. \label{eq:acf_density}
\eey 
Although this derivation was already presented in Ref.~\cite{Blazek:2011}, we repeat it because we extend the analysis to the IA statistics for the velocity field in the next subsection. Note that the conventional correlation function, 
$\xi_{\delta_A\delta_B}(r)$, is obtained by taking the average over $\theta$,
\bey
\xi_{\delta_A\delta_B}(r)=\frac{2}{\pi}\int^{\pi/2}_{0} d\theta \xi_{\delta_A\delta_B}(\vecr,\theta), \label{eq:conv_xi}
\eey
where the the anisotropic term is integrated to zero. Also, the GI correlation function, i.e., $\xi_{\delta_A +}$ defined at Eq.~(\ref{eq:gi}), is related to the alignment correlation function through 
\be
\xi_{\delta_A+}(\vecr)=\frac{2}{\pi}\int^{\pi/2}_{0} d\theta \cos(2\theta)\xi_{\delta_A\delta_B}(\vecr,\theta). 
\ee
where $b/a=0$ \cite{Okumura:2009a,Faltenbacher:2009}. While the GI and alignment correlation functions are complementary to each other, we will primarily focus on the latter because it provides direct insight into how the matter is distributed along and perpendicular to the major axis of halos.

\subsection{Velocity statistics}
\label{subsec:velocity_statistics}
In analogy to the density statistic, we consider the alignment statistics of halos/galaxies (i.e., object $B$) relative to the cosmic velocity of object $A$, $\vv_A$. Because in observations we can measure velocities along the observer's line-of-sight direction, $v_A(\vx)=\vv_A(\vx)\cdot \hat{\vx}$, we consider only the line-of-sight component of the velocity field throughout this paper. We thus use the same symbol, $v_A$, to describe the line-of-sight component of the three-dimensional velocity $\vv_A$ (see Fig.~\ref{fig:ia}). 

\subsubsection{Velocity-intrinsic ellipticity correlation}
We first introduce the velocity statistic corresponding to the GI correlation, namely the density-weighted VI correlation,
\bey
\xi_{p_A+}(\vecr)&=&\left\langle p_A(\vx_1) \wt{\gamma}_+ (\vx_2) \right \rangle \nn\\
&=&\left\langle[1+\delta_A(\vx_1)][1+\delta_B(\vx_2)]
\right. \nn 
 \\ 
&& \ \ \ \ \ \ \ \ \ \ \ \ \ \ \ \ \ \times \left.
 v_A(\vx_1)\gamma_+ (\vx_2) \right \rangle, \label{eq:vi}
\eey
where $p_A$ denotes the line-of-sight component of the density-weighted velocity, that is, the momentum field \cite{Park:2000},  
\be
p_A(\vx)\equiv \vp_A(\vx)\cdot \hat{\vx}=[1+\delta_A(\vx)]v_A(\vx), \label{eq:momentum_field}
\ee
and 
the density-weighted intrinsic ellipticity is denoted by
$\wt{\gamma}_{+}(\vx)=[1+\delta_A(\vx)]\,\gamma_{+}(\vx)$.

\subsubsection{Alignment density-momentum correlation}
In order to define the alignment velocity statistics corresponding to the alignment density correlation, we consider the two velocity statistics. The first one is the cross correlation function between momentum and density \cite{Fisher:1995}, 
$\xi_{p_A\delta_B}(\vecr)= \left\langle \left[1+\delta_A ({\bf x}_1)\right]  \left[1+\delta_B ({\bf x}_2)\right] v_{A} ({\bf x}_1)\right\rangle$. 
To define the alignment momentum-density cross correlation function, in analogy to the case of the alignment density-density correlation, we cross correlate the momentum and mass (number) density fields, using the conditional average given in Eq.~(\ref{eq:operator}): 
\bey
\left\langle p_A,\, \rho_B | \theta \right\rangle &=& \bar{\rho}_B  \left\langle \left(1+\delta_A\right)v_A,\, \left(1+\delta_B\right) | \theta
\right\rangle \nn \\
&\equiv&  \bar{\rho}_B \, \xi_{p_A\delta_B}(\vecr,\theta). \label{eq:momentum_density}
\eey
Hence, we obtain the expression for the alignment momentum-density correlation function as
\bey
\xi_{p_A\delta_B}(\vecr,\theta)&=& \langle v_A, \,1 |\theta\rangle + \langle v_A,\, \delta_B|\theta\rangle+ \langle \delta_Av_A, \,1|\theta\rangle \nn
\\
&& +  \langle \delta_Av_A,\, \delta_B|\theta  \rangle
\eey
Similarly, the alignment density-momentum correlation function is given by
\bey
\xi_{\delta_Ap_B}(\vecr,\theta)&=&\left\langle \left(1+\delta_A\right), \, \left(1+\delta_B\right) v_{B}|\theta \right\rangle \nn \\
&=& \langle \delta_A,\, v_B |\theta  \rangle + \langle \delta_A, \,\delta_B v_B |\theta  \rangle. \label{eq:density_momentum}
\eey

Using these correlation statistics, one can also define a more observationally related pairwise velocity statistic \cite{Peebles:1980,Sheth:2001,Sheth:2001a}, namely the alignment pairwise infall momentum, 
\bey
p_{AB}(\vecr,\theta)&\equiv &\xi_{\delta_Ap_B}(\vecr,\theta) - \xi_{p_A\delta_B}(\vecr,\theta) \nn\\
&=& \left\langle \left[1+\delta_A(\vx_1)\right]\left[1+\delta_B(\vx_2)\right]\right.\nn\\
&&\qquad\times\left.\left[v_B(\vx_2)-v_A(\vx_1)\right] | \theta\right\rangle .
\label{eq:infall_momentum}
\eey
Note that the subscripts $p_A$ and $p_B$ on the right-hand side are the momentum fields defined in Eq.~(\ref{eq:momentum_field}), and should not be confused with $p_{AB}$. 

\subsubsection{Alignment momentum-momentum correlation}

Another interesting velocity statistic is the momentum correlation function \cite{Gorski:1988}, given by $\xi_{p_Ap_B}(\vecr)=\left\langle \left[1+\delta_A ({\bf x}_1)\right]\left[1+\delta_B ({\bf x}_2)\right] v_{A} ({\bf x}_1)v_{B} ({\bf x}_2)\right\rangle$. The corresponding IA statistic, which we call the alignment momentum correlation function, is obtained from this quantity by replacing the standard ensemble with the conditional average in Eq.~(\ref{eq:operator}): 
\bey
\xi_{p_Ap_B}(\vecr,\theta)&=& \left\langle p_A,\, p_B | \theta \right\rangle \nn\\
&=& \left\langle \left(1+\delta_A\right)v_A, \,\left(1+\delta_B\right)v_B | \theta \right\rangle \nn\\
&=& \langle v_A,\, v_B|\theta \rangle+\langle \delta_Av_A,\, v_B|\theta \rangle  \nn \\
 &&+\langle v_A,\, \delta_Bv_B|\theta \rangle+ \langle \delta_Av_A,\, \delta_Bv_B|\theta \rangle. \label{eq:momentum_momentum}
\eey
To make a clear distinction from similar quantities, we shall denote it by $\psi_{AB}(\vecr,\theta)\equiv \xi_{p_Ap_B}(\vecr,\theta)$ hereafter.

Finally, we introduce the pair-weighted velocity dispersion, $\Sigma_{AB}^2(\vecr)$, and apply the orientation-dependent average to it. The alignment pairwise velocity dispersion then becomes
$\Sigma_{AB}^2(\vecr,\theta)$, and it is expressed as
\bey
\Sigma_{AB}^2(\vecr,\theta)&=&\Bigl\langle \left[1+\delta_A(\vx_1)\right] \left[1+\delta_B(\vx_2)\right] \nn \\
&& \qquad\qquad \times \left[v_{B} ({\bf x}_2)-v_{A} ({\bf x}_1)\right]^2|\theta\Bigr\rangle \nn \\
&=& \langle v_A^2,\,1 | \theta \rangle + \sigma_{v_B}^2 - 2\psi_{AB}(\vecr,\theta) +\langle \delta_B v_B^2 \rangle \nn \\
&&+\langle \delta_{A}, \,v_B^2 | \theta \rangle+\langle v_A^2, \,\delta_{B} | \theta \rangle +\langle \delta_{A}v_A^2, \,1 | \theta \rangle  \nn \\
&&+\langle \delta_{A}v_A^2, \,\delta_{B} | \theta \rangle +\langle \delta_{A},\, \delta_{B}v_B^2 | \theta \rangle , \label{eq:velocity_disp}
\eey
where $\sigma_{v_B}$ is the one-dimensional velocity dispersion defined by $\sigma_{v_B}^2 = \left\langle v_B^2 \right\rangle$

Similarly to the density correlation function, the conventional velocity statistics, $p_{AB}(\vecr)$, $\psi_{AB}(\vecr)$, and $\Sigma_{AB}^2(\vecr)$, can be obtained by averaging over $\theta$,
\bey
X_{AB}(\vecr)=\frac{2}{\pi}\int^{\pi/2}_{0} d\theta X_{AB}(\vecr,\theta), 
\eey
with $X$ being $p$, $\psi$ or $\Sigma^2$.

\section{Theoretical predictions of IA density and velocity statistics}\label{sec:theoretical_prediction}

\subsection{Conditional average of Gaussian random fields}\label{sec:gaussian_random}

Since we are interested in the statistical properties at very large scales, $r\gg 10\himpc$, predictions based on linear theory calculations basically give an accurate description, and the assumption that all the relevant fields follow the Gaussian statistics is validated. In such a case, all the statistical quantities can be computed with multivariate Gaussian distribution, with the covariance matrix estimated from linear theory \cite{Bardeen:1986}.   

Consider a set of Gaussian random fields, $\vq^t=(q_1,q_2,\cdots, q_N)$, with zero expectation value, $\langle \vq \rangle = 0$. An ensemble average of the quantity $F(\vq)$ is then expressed as
\bey
&& \langle F \rangle = \int dq_1dq_2\cdots dq_N \nn \\
&& \ \ \ \ \ \ \ \ \ \ \ \  \times \frac{F(\vq)}{(2\pi)^{N/2}|\det{\bf C}|^{1/2}} e^{-(1/2)\,\vq^t \vC^{-1}\vq }, \label{eq:gaussian}
\eey
where $\vC$ is the covariance matrix defined by $\vC=\left\langle \vq\vq^t\right\rangle$. For the IA statistics considered in the previous section, a relevant set of the fields $\vq$ is 
\be
\vq^t= (\delta_A,\delta_B,v_A,v_B,\gamma_+,\gamma_\times). \label{eq:vector_q}
\ee
Equation (\ref{eq:gaussian}) with Eq.~(\ref{eq:vector_q}) provides a basis to derive analytical expressions for the statistical quantities in Sec.~\ref{sec:theory}. The calculation with Eq.~(\ref{eq:vector_q}) is an extension of the work by Refs.~\cite{Fisher:1995,Blazek:2011} \footnote{Reference \cite{Fisher:1995} took the vector $\vq$ as $\vq^t= (\delta_A,\delta_B,v_A,v_B)$, while in Ref. \cite{Blazek:2011}, $\vq$ was considered to be $\vq^t= (\delta_A,\delta_B,\gamma_+,\gamma_\times)$. }. Note that the intrinsic ellipticity characterized by $\gamma_+$ and $\gamma_\times$ is always defined at the location of the tracer $B$, namely at $\vx_2$, and the two velocities, $v_A$ and $v_B$, are measured along the line of sight. Here the ellipticities, $\gamma_+$ and $\gamma_\times$, are assumed to be Gaussian random fields. 
The model we consider in order to relate the ellipticities to the underlying matter field is compatible with the assumption of Gaussianity (see Sec.~\ref{sec:la_model}).

To derive analytical expressions for the alignment statistics, we follow Ref. \cite{Blazek:2011}, and switch the integration variables from $(q_5,q_6)=(\gamma_+,\gamma_\times)$ to $(q_5',q_6')=(\gamma_0,\theta)$ through Eq.~(\ref{eq:gamma_pc}). Then, any alignment statistic, expressed in terms of Eq.~(\ref{eq:operator}), can be computed by integrating over five variables, while keeping $\theta$ fixed \footnote{Unlike in the previous section, we need to define $\theta$ at any location $\vx_2$, irrespective of the presence or absence of the tracer $B$ at that location in practice to perform actual calculations. This difference is because the analytical formulation here is based on the Eulerian perturbation theory (or volume-weighted quantities as the basic building blocks), while the final statistics are always density-weighted quantities defined through discrete tracers. Here, we employ a specific IA model to describe $\gamma$ (or $\theta$) at an arbitrary spatial coordinate. After summing up the relevant volume-weighted quantities to express density-weighted ones, again, the ambiguity in the direction $\theta$ at locations without a tracer does not enter the final results.}. To be explicit, using $d\gamma_+d\gamma_\times = 2\gamma_0d\gamma_0d\theta$, the conditional average in Eq.~(\ref{eq:operator}) is expressed as
\begin{align}
\langle X,\,\, Y|\theta \rangle &= \int d\delta_A d\delta_B dv_A dv_B d\gamma_0\,
\nonumber
\\
&\times
\frac{\gamma_0}{(2\pi)^2} e^{-(1/2)\,\vq^t \vC^{-1}\vq }\,X\,Y.
\label{eq:conditional_average_integral}
\end{align}

In Appendix \ref{sec:formula}, we present several useful formulas for the integral given above, which are used to derive analytical expressions for the IA statistics. The last step to derive the results requires an explicit form of the covariance matrix, $\vC$. Our $6\times 6$ covariance matrix for the vector $\vq$ [Eq.~(\ref{eq:vector_q})] is explicitly given by
\bey
\vC=\left(
\begin{array}{cccccc}
\sigma_{\delta_A}^2 & \xi_{\delta_A\delta_B} & 0 & \xi_{\delta_A v_B} & \xi_{\delta_A+} & 0 \\
\xi_{\delta_A\delta_B} & \sigma_{\delta_B}^2 & \xi_{v_A \delta_B} & 0 & 0 & 0 \\
0 & \xi_{v_A\delta_B} & \sigma_{v_A}^2 & \xi_{v_Av_B} & \xi_{v_A+} & 0 \\
\xi_{\delta_A v_B} & 0 & \xi_{v_Av_B} & \sigma_{v_B}^2 & 0 & 0 \\
\xi_{\delta_A+} & 0 & \xi_{v_A +} & 0 & \sigma_{\gamma}^2 & 0 \\
0 & 0 & 0 & 0 & 0 & \sigma_{\gamma}^2 \\
\end{array}
\right), \label{eq:covariance}
\eey
where the diagonal components (variances) are,
$\sigma_{\delta_A}^2= \left\langle \delta_A^2\right\rangle$, 
$\sigma_{\delta_B}^2= \left\langle \delta_B^2\right\rangle$, 
$\sigma_\gamma^2= \left\langle \gamma_+^2\right\rangle= \left\langle \gamma_\times^2\right\rangle$, $\sigma_{v_A}^2= \left\langle v_A^2\right\rangle$, and $\sigma_{v_B}^2= \left\langle v_B^2\right\rangle$. In the absence of velocity bias, we generally have $\sigma_{v_A}^2=\sigma_{v_B}^2$ in linear theory. 
The term $\xi_{\delta_A v_B}(\vecr)$ is the cross-correlation function between the density of field $A$ and the line-of-sight velocity of field $B$ and $\xi_{v_Av_B}(\vecr)$ is the velocity correlation function of velocity fields $A$ and $B$. The term $\xi_{\delta_A + }(\vecr)$ is the cross-correlation function between the density field $A$ and the ellipticity of the field $B$,  $\xi_{\delta_A + }(\vecr) = \left\langle \delta_A(\vx_1)\gamma_+(\vx_2)\right\rangle$ and $\xi_{v_A+}(\vecr)=\left\langle v_A(\vx_1)\gamma_+(\vx_2)\right\rangle$ is the velocity-intrinsic ellipticity correlation function. Note that all the fields in the covariance matrix including ellipticities are defined as volume-weighted quantities. All the elements will be computed later in Secs. \ref{sec:la_model} and \ref{sec:linear_pt}.

\subsection{Analytical expressions for IA density and velocity statistics}\label{sec:prediction}

Here, we summarize the analytical expressions for the IA statistics introduced in Sec.~\ref{sec:theory_density}. 

Consider first the alignment density correlation, $\xi_{\delta_A\delta_B}$. Letting $F=\delta_A+\delta_A\delta_B$ and using Eqs. (\ref{eq:gaussian_new_integral}), (\ref{eq:formula1}) and (\ref{eq:formula2}) with the given fields being Gaussian, the following expression is derived from Eq.~(\ref{eq:acf_density}) straightforwardly:
\bey
\xi_{\delta_A\delta_B}(\vecr,\theta)&=&\xi_{\delta_A\delta_B}(r) \nn \\ 
&+& \sqrt{\frac{\pi}{2\sigma_\gamma^2}}\xi_{\delta_A+}(\vecr)\cos{(2\theta)}, \label{eq:acf_density_formula}
\eey
where the first and second terms are respectively derived from the first and second terms in Eq. (\ref{eq:acf_density}). The expression of this statistic has already been derived by Ref.~\cite{Blazek:2011}, and we have rederived it using the formula presented in Appendix \ref{sec:formula}. 

Next consider the pairwise infall momentum $p_{AB}$, given in Eq.~(\ref{eq:infall_momentum}). The term $\langle \delta_A,\,\delta_B v_B |\theta  \rangle$ vanishes, and other terms in $\xi_{\delta_Ap_B}$ and $\xi_{p_A\delta_B}$ can be computed with the help of formulas given in Eqs. (\ref{eq:formula1})--(\ref{eq:formula3}). The resultant expression becomes
\bey
p_{AB}(\vecr,\theta)
&=& \xi_{\delta_A v_B}(\vecr)-\xi_{v_A\delta_B}(\vecr) -\sqrt{\frac{\pi}{2\sigma_\gamma^2}}\cos{(2\theta)} \nn \\
&& \times \left\{ [1+\xi_{\delta_A\delta_B}(\vecr)]\xi_{v_A+}(\vecr) +\xi_{v_A\delta_B}(\vecr)\xi_{\delta_A +}(\vecr) \right\} \nn \\
&& - \frac{1}{\sigma_\gamma^2} \cos{(4\theta)}\xi_{v_A+}(\vecr)\xi_{\delta_A +}(\vecr). \label{eq:acf_infall_formula}
\eey
The expression includes not only the terms proportional to $\cos{(2\theta)}$, but also the higher-order contributions with $\cos{(4\theta)}$.

Similarly, the alignment momentum correlation $\psi_{AB}$ is also obtained, using the formulas in Appendix \ref{sec:formula} and the fact that the term $\langle \delta_Bv_Av_B|\theta \rangle$ vanishes. The final form of the expression becomes
\bey
\psi_{AB}(\vecr,\theta)&=& [1+\xi_{\delta_A\delta_B}(\vecr)]\xi_{v_A v_B }(\vecr)+\xi_{\delta_A v_B}(\vecr)\xi_{v_A\delta_B}(\vecr)  \nn \\
&+&\sqrt{\frac{\pi}{2\sigma_{\gamma}^2}}\cos{(2\theta)} \left[ \xi_{\delta_A v_B }(\vecr)\xi_{v_A  +}(\vecr) \right. \nn \\
&& \ \ \ \ \ \ \ \ \ \ \ \ \ \ \ \ \ \ \ \ \ \left .+ \xi_{v_Av_B}(\vecr)\xi_{\delta_A +}(\vecr)\right]
\label{eq:acf_momentum_formula}
\eey
Unlike the pairwise infall momentum, in the momentum correlation the alignment-dependent terms proportional to $\cos(2\theta)$ are given by the products of two correlation functions, $\xi_{\delta_A v_B }\xi_{v_A  +}$ and $\xi_{v_Av_B}\xi_{\delta_A +}$. This implies that the alignment dependence will disappear at large scales.

Finally, the density-weighted pairwise velocity dispersion can be expressed under the assumption of the Gaussian random fields as 
\begin{widetext}
\bey
\Sigma_{AB}^2(\vecr,\theta) 
 &=& [1+\xi_{\delta_A\delta_B}(\vecr)][\sigma_{v_A }^2 +\sigma_{v_B }^2 - 2\xi_{v_A v_B }(\vecr)] 
 -2 \xi_{\delta_A v_B}(\vecr)\xi_{v_A\delta_B}(\vecr)
 \nn \\
&+& \sqrt{\frac{\pi}{2\sigma_\gamma^2}}\left\{ 
[\sigma_{v_A}^2+\sigma_{v_B}^2 
- 2\xi_{v_Av_B}(\vecr)]\xi_{\delta_A +}(\vecr)  
-2[\xi_{\delta_A v_B }(\vecr)
-\xi_{v_A\delta_B}(\vecr)]\xi_{v_A  +}(\vecr)          
- \frac{3}{4\sigma_\gamma^2} \xi_{\delta_A+}(\vecr)\xi_{v_A+}^2(\vecr) 
\right\}  \cos{(2\theta)}  \nn \\
&+& \frac{1}{\sigma_\gamma^2}\left\{   [1+\xi_{\delta_A\delta_B}(\vecr)] \xi_{v_A+}(\vecr)
+ 2 \xi_{\delta_A+}(\vecr)\xi_{v_A\delta_B}(\vecr) \right\}\xi_{v_A+}(\vecr)\cos{(4\theta)} 
\nn \\ &+& 
\frac{3}{4\sigma_{\gamma}^2}\sqrt{\frac{\pi}{2\sigma_\gamma^2}}\xi_{\delta_A+}(\vecr)\xi_{v_A+}^2(\vecr)\cos{(6\theta)}. \label{eq:velocity_disp_formula}
\eey 
\end{widetext}
The resultant expression includes the alignment-dependent terms proportional to $\cos{(2\theta)}$, $\cos{(4\theta)}$ and $\cos{(6\theta)}$, which are all expressed as products of multiple correlation functions. Thus, the orientation dependence becomes negligible at large scales, and it would be important only at small scales. 

\subsection{Linear alignment model}\label{sec:la_model}

For a quantitative calculation of the analytical expressions in Sec.~\ref{sec:prediction}, the covariance given in Eq.~(\ref{eq:covariance}) has to be estimated. In doing so, we need a model of intrinsic ellipticity, which relates $\gamma_{(+,\times)}$ to the density or velocity field. In this paper, we adopt the LA model which is one of the simplest models to describe the ellipticity/orientation of elliptical galaxies or halos. In the LA model, the intrinsic ellipticity [Eq.~(\ref{eq:gamma_pc})] is assumed to follow a linear relation with the Newtonian potential, $\Psi_P$
\cite{Catelan:2001,Hirata:2004},
\be
\gamma_{(+,\times)}(\vx)= -\frac{C_1}{4\pi G}\left( \nabla_x^2-\nabla_y^2, 2 \nabla_x\nabla_y \right) S[\Psi_P], 
\label{eq:gamma_la}
\ee
where $G$ is the Newtonian gravitational constant, and $C_1$ parametrizes the strength of the IA. The function $S$ represents a smoothing filter that introduces a cutoff to the fluctuations on halo scales. The $x$ and $y$ axes are taken to be on the plane of the sky, and thus the $z$ axis indicates the line-of-sight direction. The potential $\Psi_P$ is related to the density field via the Poisson equation. In Fourier space, we have
\be
\Psi_P(\vk) = -4\pi G\frac{\bar{\rho}(z)}{\bar{D}(z)} a^2k^{-2} \delta(\vk), \label{eq:poisson}
\ee
with $\bar{\rho}$ being the mean density of the Universe. The function is  $\bar{D}(z)\propto (1+z)D(z)$, where $D(z)$ is the linear growth factor. 

Using Eqs.~(\ref{eq:gamma_la}) and (\ref{eq:poisson}), the Fourier transform of the density-weighted intrinsic ellipticity, $\wt{\gamma}_{(+,\times)}(\vx)$, is described as,
\bey
\wt{\gamma}_{(+,\times)} (\vk) &=&  \frac{-C_1\bar{\rho}(z)}{\bar{D}(z)}a^2 \int d^3 \vk' d^3 \vk'' 
\delta_D(\vk-\vk'-\vk'') \nn \\
&&\times \left(  {k'}_{x}^{2}- {k'}_{y}^{2}, 2{k'}_x{k'}_y \right)  {k'}^{-2} \delta(\vk') \nn \\
&&\times \left [ (2\pi)^3 \delta_D(\vk'') + \delta (\vk'') \right],
\eey
where $\delta_D(\vk)$ is the Dirac delta function. With this expression, the linear theory estimate of the cross-correlation function between the density field and the ellipticity is given by
\bey
\xi_{\delta_A + }(\vecr) 
&=&\frac{C_1 \bar{\rho}}{\bar{D}}a^2 b_A \int^{\infty}_{0} \frac{k_\perp dk_\perp}{2\pi^2} J_2(k_\perp r_\perp) 
\nn \\
&& \times \int^{\infty}_0 dk_{\parallel} \frac{k_\perp^2}{k^2} P_{\delta\delta}(k) \cos{(k_\parallel r_\parallel)} ,
\label{eq:gi_la}
\eey
where $k_\perp^2=k_x^2+k_y^2$ and $k_\parallel = k_z$ (hence, $k^2=k_\perp^2+k_\parallel^2$), and the quantities $r_\perp$ and $r_\parallel$ are the separation perpendicular and parallel to the line-of-sight direction ($r^2=r_\perp^2+r_\parallel^2$). The function $J_2$ is the second-order Bessel function and $b_A$ is the linear bias parameter for objects $A$. Likewise, the velocity-ellipticity correlation at linear order is expressed as
\bey
\xi_{v_A+}(\vecr) &=& \frac{C_1 \bar{\rho}}{\bar{D}} a^3fH \int^{\infty}_{0} \frac{k_\perp dk_\perp}{2\pi^2} J_2(k_\perp r_\perp)
\nn \\
&&\times \int^{\infty}_0 dk_{\parallel} \frac{k_\perp^2 k_\parallel}{k^4}
P_{\delta\Theta}(k)\sin{(k_\parallel r_\parallel)}, \label{eq:vi_la}
\eey
where $\Theta$ is the velocity-divergence field, defined by $\Theta(\vx)=-\nabla \cdot \vv/(aHf)$. The function $H(a)$ is the Hubble parameter, and $f$ is the linear growth rate, given by $f\equiv d\ln D/d\ln a$. 
Here and in the next subsection, the three power spectra are introduced, $P_{\delta\delta}$, $P_{\Theta\Theta}$ and $P_{\delta\Theta}$, which respectively denote the auto power spectra of the density, velocity divergence, and their cross spectrum. 

\subsection{Covariance in linear theory }\label{sec:linear_pt}

Based on linear theory, the remaining quantities to calculate the covariance are analytically expressed. Assuming the linear bias relation between objects of our interest and mass-density fluctuations, the density-density, density-velocity, and velocity-velocity correlation functions become 
\bey
\xi_{\delta_A\delta_B}(r) &=& b_Ab_B \int \frac{k^2dk}{2\pi^2}P_{\delta\delta}(k)j_0(kr)  , \label{eq:xi_dd}\\
\xi_{\delta_A v_{B} }(\vecr) &=& -aHf b_A  \mu \int \frac{kdk}{2\pi^2} P_{\delta\Theta}(k)j_1(kr) , \label{eq:xi_dt} \\
\xi_{v_A v_B }(\vecr) &=& (aHf)^2 \left[ \frac{1}{3}\int\frac{dk}{2\pi^2}P_{\Theta\Theta}(k)j_0(kr) \right. \nn \\
&&  \left. +\left( \frac{1}{3}-\mu^2 \right) \int\frac{dk}{2\pi^2}P_{\Theta\Theta}(k)j_2(kr) \right] , \label{eq:xi_tt}
\eey
where $j_\ell$ is the spherical Bessel function. 
Equations (\ref{eq:xi_dt}) and (\ref{eq:xi_tt}) have explicit directional dependence described by $\mu$, which is the direction cosine between the line-of-sight direction and the separation vector $\vecr$, i.e., $\mu=\hat{\vx}\cdot\hat{\vecr}=r_\parallel / r$. Note that the angle $\cos^{-1}{(\mu)}$ and the orientation of the halo major axis $\theta$ are completely different quantities, and should not be confused (see Fig.~\ref{fig:ia}). As it is obvious from Eqs.~(\ref{eq:xi_dt}) and (\ref{eq:xi_tt}), if we expand the expressions in Legendre polynomials $\mathcal{P}_\ell(\mu)$, the function $\xi_{\delta_Av_B}$ has only a dipole moment $(\ell=1)$, while $\xi_{v_Av_B}$ has both monopole $(\ell=0)$ and quadrupole $(\ell=2)$ moments. 

Finally, the velocity dispersion is described in linear theory as
\begin{align}
\sigma_{v_A}^2 &= \xi_{v_Av_A}(0)=(aHf)^2 \int \frac{dk}{6\pi^2}P_{\Theta\Theta}(k)
\nonumber\\
&=\sigma_{v_B}^2. \label{eq:vel_var} 
\end{align}

Provided the power spectra $P_{\delta\delta}$, $P_{\delta\Theta}$, and $P_{\Theta\Theta}$, all the expressions summarized in Secs.~\ref{sec:la_model} and \ref{sec:linear_pt}, i.e., Eqs.~(\ref{eq:gi_la})--(\ref{eq:vel_var}), can be computed. Note that in the linear theory limit, we have $P_{\Theta\Theta} = P_{\delta\Theta}=P_{\delta\delta}$. Below, we will use the linear power spectrum obtained from \camb code \cite{Lewis:2000}. As a simple nonlinear extension of the LA model, we may replace the linear power spectrum with its nonlinear counterpart \cite{Bernardeau:2002}. This is called the nonlinear alignment (NLA) model \cite{Bridle:2007,Blazek:2011,Chisari:2013}. Since we are interested in the IA statistics at large scales, we shall mainly use the LA model. The NLA model is considered only when we focus on the BAO scales, and we use the \regpt code to compute the three nonlinear power spectra, $P_{\delta\delta}$, $P_{\delta\Theta}$, and $P_{\Theta\Theta}$ \cite{Taruya:2012,Taruya:2013}. This code gives the perturbation theory (PT) predictions based on a resummed perturbative calculation, which is basically applicable in the weakly nonlinear regime.  By comparing with simulations, we will show how well the nonlinear damping of the BAO feature in different statistics can be modeled with this code.


\section{Analysis}\label{sec:analysis}
\subsection{$N$-body simulations and subhalos}\label{sec:nbody}\
In order to study the alignment statistics, we use a series of large and high-resolution $N$-body simulations of the $\Lambda$CDM cosmology seeded with Gaussian initial conditions. These are performed as a part of the \darkquest project \cite{Nishimichi:2018}. We here use outputs of the low-resolution runs performed to explore large-scale correlation signals for massive halos. We employ $n_p=2048^3$ particles of mass $m_p= 8.15875\times 10^{10}\himsun$ in a cubic box of side $L_{\rm box} = 2\higpc$. In total, eight independent realizations are simulated and the snapshots at $z=0.306$ are used. 

Subhalos are identified using phase-space information of matter particles, the   \rockstar algorithm \cite{Behroozi:2013}. The velocity of the (sub)halo is determined by the average particle velocity within the innermost 10\% of the (sub)halo radius.  We use the standard definition for the halo radius and mass of
\be
M_h \equiv M_{\rm \Delta m} = M(<R_{\rm\Delta m}) = (4\pi/3)\Delta \rho_{\rm m}(z) R_{\rm\Delta m}^3,
\ee
where $\rho_{\rm m}$ is the mean mass density of the Universe at a given redshift $z$, and we adopt $\Delta = 200$. The main \rockstar output does not distinguish between halos and subhalos. We thus utilize a separate routine provided as a part of the \rockstar package to make the distinction using the radius $R_{\rm 200 m}$ as the boundary. If a halo center is inside $R_{\rm 200m}$ of a neighboring more massive halo, we classify it as a satellite subhalo. If two or more subhalos are located within the virial radius of each other, the most massive one is labeled as a central subhalo. 

To study cluster-scale halos, we select subhalos with $M_{h} \geq 10^{14}h^{-1}M_\odot$, which roughly corresponds to the typical threshold of the richness parameters used by the cluster-finding algorithms in the literature. We label these massive halos as ``clusters.'' We will present the analysis of the IAs of the cluster shapes relative to the density and velocity fields traced by dark matter/galaxies/clusters. To this end, we create mock galaxy catalogs using a halo occupation distribution (HOD) model \cite{Zheng:2005} applied for the LOWZ galaxy sample of the SDSS-III Baryon Oscillation Spectroscopic Survey obtained by Ref.~\cite{Parejko:2013}. We populate halos with galaxies according to the best-fitting HOD $N(M_h)$. For central subhalos that contain satellite galaxies, besides the halo center chosen as a central galaxy, we randomly draw $N(M_h)-1$ member subhalos to mimic the positions and velocities of the satellites (see Refs.~\cite{Nishimichi:2014,Okumura:2017} for alternative methods). We use a random selection of subhalos rather than the largest subhalos because a satellite subhalo undergoes tidal disruption in the host halo and its mass decreases as it goes toward the center of the gravitational potential. We assume halos to have triaxial shapes \cite{Jing:2002a} and estimate the orientations of their major axes using the second moments of the distribution of member particles projected onto the celestial plane. Table \ref{tab:halo} summarizes properties of our mock samples.

\begin{table}[b]
\caption{Properties of mock subhalo samples at $z=0.306$. $f_{sat}$ is the number fraction of satellites, $M_\mathrm{min}$ and $\overline{M}$ are the minimum and average halo mass in units of $10^{12}h^{-1}M_\odot$, respectively,  
$\overline{n}$ is the number density in units of $h^3{\rm Mpc}^{-3}$,
and $b_A$ ($A=\{c,g\}$) is the cluster/galaxy bias computed in the large-scale limit.}
\begin{center}
\begin{tabular}{lcccccc}
\hline\hline
Types & Label &$f_{sat}$ & $M_{\rm min}$ &   $\overline{n}$ &$b_A$ & $\overline{M}$ \\ 
\noalign{\hrule height 1pt}
Clusters & $c$ &0 &100 & $ 2.05 \times 10^{-5}$ & $3.11$ & 188\\
Galaxies  & $g$ &0.137 & 1.63 &   $5.27\times 10^{-4}$ &$1.70$ & 25.2 \\ 
\hline\hline
\end{tabular}
\end{center}
\label{tab:halo}
\end{table}

Note that, due to the limited hard disk space, the information of dark matter particles could have been stored partially and thus was lost for four realizations out of eight \textit{after} the measurement of the mass moments necessary to determine the direction of the major axis. Hence, we could not measure some of the statistics for which the information of dark matter particles is needed, while the information of the halos including the direction of the major axis traced by the dark matter particles was available for all eight realizations. Thus, if presented statistics include the density or velocity field of dark matter in the following analysis, the result is obtained from four realizations; otherwise it is out of the entire eight realizations. 

The linear bias parameter of sample $A$ is determined by measuring $b_A(r)=[\xi_{\delta_A\delta_A}(r)/\xi_{\delta_m\delta_m}(r)] ^{1/2}$ and searching for the best-fitting constant over $20\himpc <r<80 \himpc$,  where $\xi_{\delta_m\delta_m}$ is the matter correlation function measured from the same simulation as the numerator. The determined bias is consistent with that obtained from the cross-correlation function, $b_A(r)=\xi_{\delta_A\delta_m}(r)/\xi_{\delta_m\delta_m}(r)$. The values of the bias are shown in Table \ref{tab:halo}. These bias parameters are used for our theoretical modeling in Sec.~\ref{sec:comparison}.

\subsection{Estimators}
Here we describe the estimators we use to measure the alignment statistics from $N$-body simulations. As mentioned in the previous section, each statistic $X$ can be expanded in the Legendre polynomials ${\cal P}_\ell(\mu)$ and it has multipole components. For the GI and VI correlations ($X=\{\xi_{\delta_A+},\xi_{p_A+}\}$) we have
\be
X_{\ell}(r)=(2\ell + 1)\int^1_0 X(r,\mu){\cal P}_\ell(\mu) d\mu,
\ee 
while for the other alignment statistics, the $\theta$ dependence needs to be additionally included. In this paper, we consider only the lowest moment for each statistic, either the monopole or dipole, and the subscript $\ell$ is omitted in the following.

The estimator of the GI correlation function between the density of sample $A$ and ellipticity of sample $B$ is given by Ref.~\cite{Mandelbaum:2006}
\be
\xi_{\delta_A+}(r)=\frac{\sum_{i,j|r}{\gamma_+(j|i)}}{ R_AR_B(r)}, 
\ee
where we consider only the monopole moment. The sum in the numerator is taken over all pairs weighted by the plus component of ellipticity in the separation $r$, where $i$ and $j$ run over samples $A$ and $B$, respectively, and $\gamma_+(j|i)$ is the $+$ component of the ellipticity of the $j$th shape for sample $B$ measured relative to the direction to the $i$th tracer for sample $A$ (see Fig.~\ref{fig:ia}). The denominator $R_AR_B(r)$ is the pair count of the the random distributions as a function of separation $r$. It can be analytically and accurately computed because we place the periodic boundary condition on the simulation box, and we will set the number density of the random sample $A$ ($B$) to be equivalent to that of the data sample $A$ ($B$) so that the pair count does not need to be normalized. 

The VI correlation contains odd-order multipoles and the lowest-order term is the dipole, because we use the line-of-sight component of the velocity, $v_A(\vx)=\vv_A(\vx)\cdot \hat{\vx}$. Hence, we propose the following estimator for the density-weighted VI correlation dipole: 
\be
\xi_{p_A+}(r)=\frac{\sum_{i,j |r} {v_A(\vx_i)\gamma_{+} (\vx_j)} \mu_{ij} }{\sum_{i,j |r} {\mu_{ij,{\rm rand}}^2}},
\ee
where $\mu_{ij}\equiv \hat{\vecr}\cdot \frac{\hat{\vx}_i+\hat{\vx}_j}{2}$ is the directional cosine between the separation vector of each pair and the line of sight, and $\mu_{ij,{\rm rand}} $ is the same as $\mu_{ij}$ but for the random distributions. Thus, the denominator is similar to $R_AR_B$ but each pair is weighted by the square of the direction cosine. In this analysis, the shape information is always taken from clusters, while the field $A$ can be either matter, galaxies or clusters themselves. 

We then present estimators for the alignment density and velocity correlation statistics. The alignment correlation function of the density ($A$) and shape ($B$) samples can be measured by
\be 
\xi_{AB}(r,\theta)=\frac{D_AD_B(r,\theta)}{R_AR_B(r,\theta)}-1,
\ee
where $D_AD_B(r,\theta)$ is the pair counts of the data as functions of separation $r$ and angle $\theta$. As is the case with the VI correlation, the dominant contribution for the alignment pairwise infall momentum is the dipole moment. We adopt an estimator for the pairwise momentum dipole \cite{Ferreira:1999,Okumura:2014}, 
\be
p_{AB}(r,\theta)=\frac{\sum_{i,j |r,\theta} {[v_A(\vx_i) -v_B(\vx_j)] \mu_{ij} }}{\sum_{i,j |r,\theta} {\mu_{ij,{\rm rand}}^2}}.
\ee
From Eq.~(\ref{eq:acf_momentum_formula}), the conventional momentum correlation function contains contributions from the monopole and quadrupole moments, and the higher-order moments are produced by IAs. Thus, here we consider only the monopole component for the alignment momentum correlation function and use the estimator
\be
\psi_{AB}(r,\theta)=\frac{\sum_{i,j |r,\theta}{v_{A}(\vx_i)v_{B}(\vx_j)}}{R_AR_B(r,\theta)},
\ee
where 
the numerator is the pair count of samples $A$ and $B$ weighted by the products of the line-of-sight components of their velocities as functions of $r$ and $\theta$.
Likewise, for the (density-weighted) alignment velocity dispersion which has the monopole moment as the dominant contribution followed by the quadrupole, we use the estimator for the monopole,
\be
\Sigma_{AB}^2(r,\theta)=\frac{\sum_{i,j |r,\theta} {[v_A(\vx_i) -v_B(\vx_j)]^2 \mu_{ij} }}{ R_AR_B(r,\theta)}.
\ee

Once again, the shape information is taken from clusters, so that sample $B$ is always a cluster, $B=\{c\}$. 
In the following analysis, we measure these statistics where $A=\{m\}$ and $A=\{g,c\}$ from four and eight realizations, respectively, and present the means and their errors from the scatters .

\subsection{Measurements}\label{sec:measurements}

\begin{figure}[bt]
\includegraphics[width=0.49\textwidth,angle=0,clip]{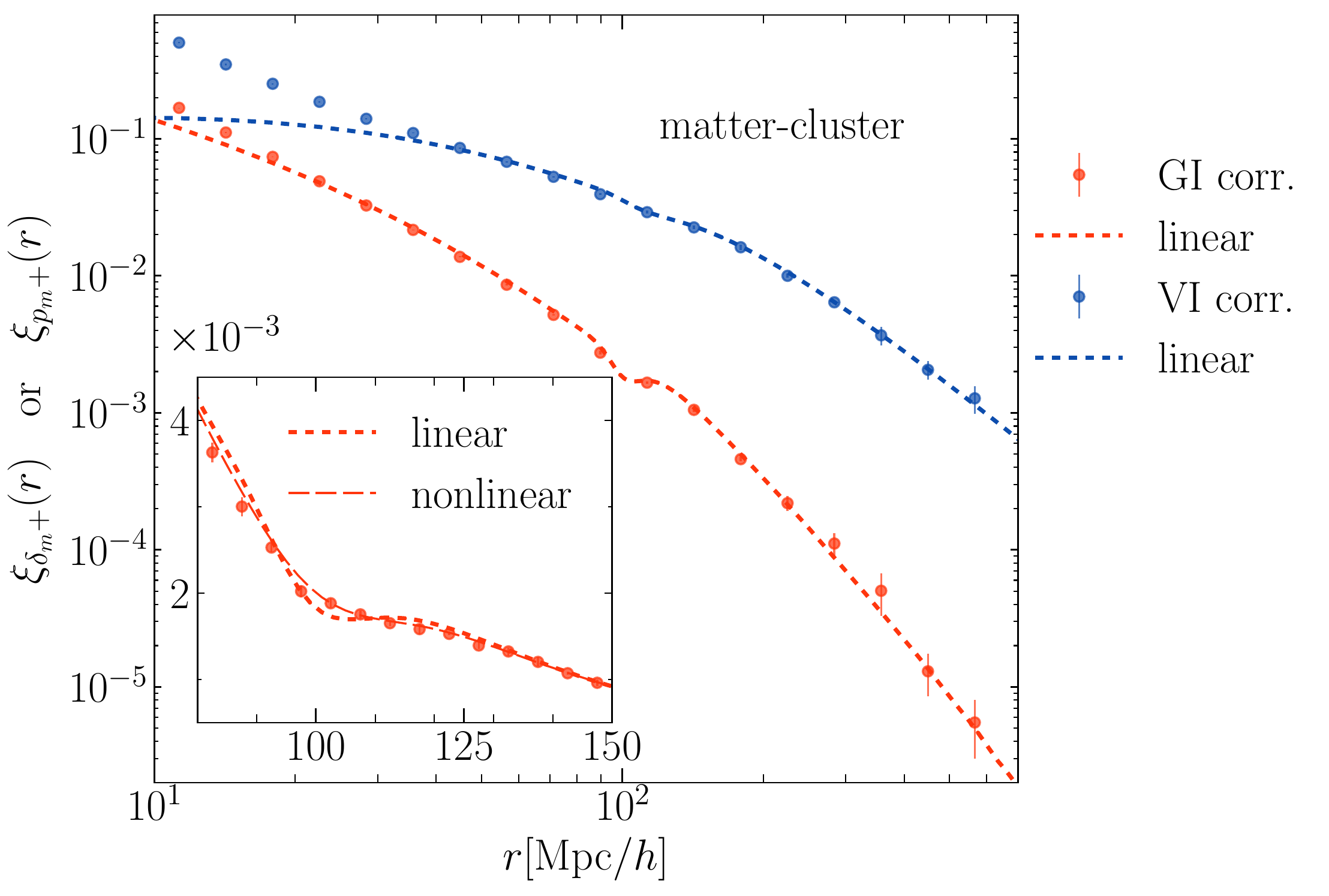}
\caption{GI and VI correlation functions shown by the red and blue points respectively. The density/velocity field is sampled by dark matter and the ellipticity field is sampled by clusters. The dotted curves are the LA model predictions. The inset shows an expanded view on BAO scales for the GI correlation with a linear vertical axis. In the inset we also show the NLA model using \regpt as the dashed curve.
}
\label{fig:gi_m_r_2gpc_z014}
\end{figure}

We start by presenting the GI correlation which has been extensively studied in the literature in the context of weak lensing systematics. In Fig.~\ref{fig:gi_m_r_2gpc_z014}, the red points show the GI correlation function of cluster shapes with dark matter, $\xi_{\delta_m+}(r)$. The inset shows the zoomed view of the GI correlation around the BAO scales. The blue points show a measurement of the density-weighted VI correlation function, $\xi_{p_m+}$. 

\begin{figure}[bt]
\includegraphics[width=0.38\textwidth,angle=0,clip]{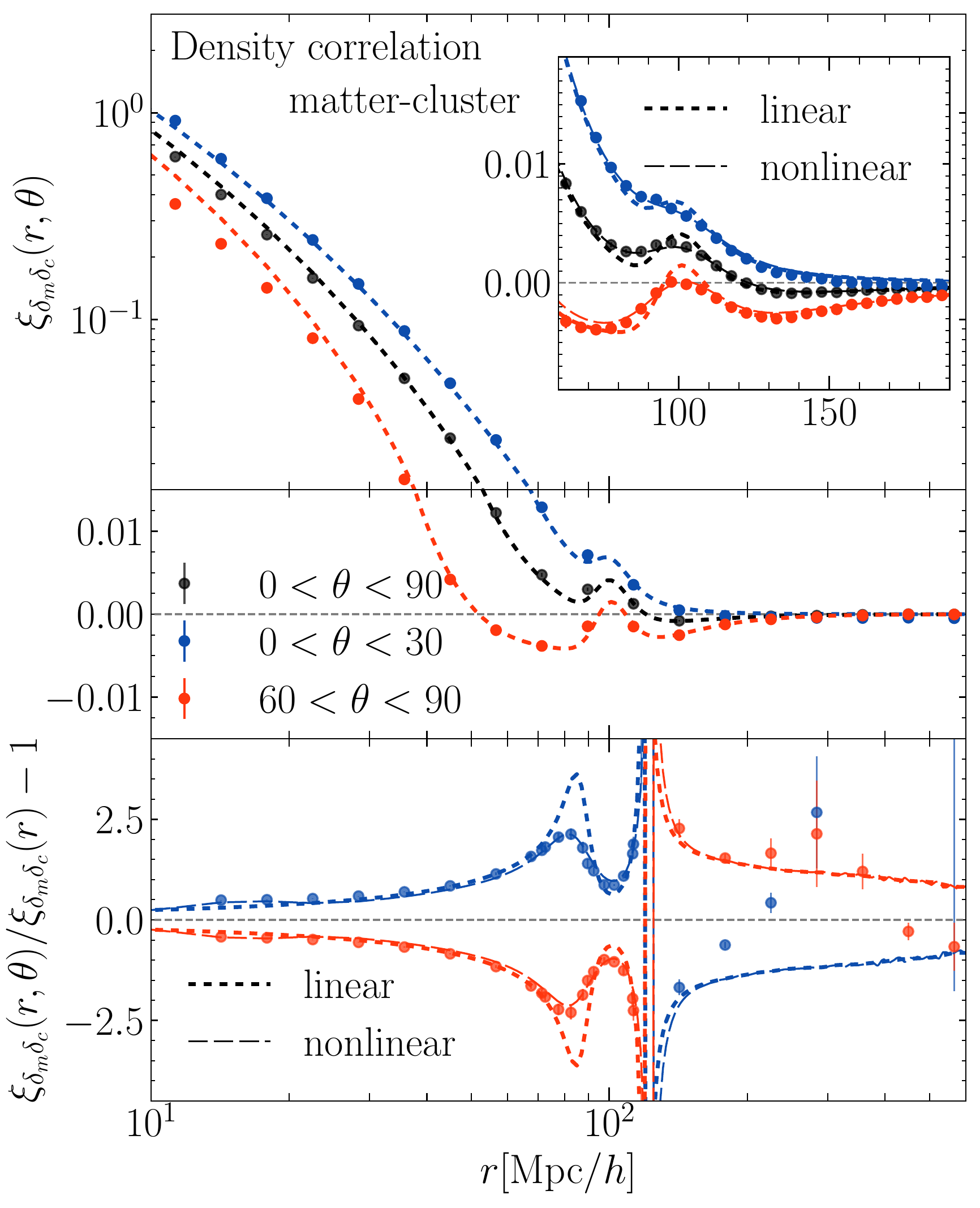}
\caption{
Alignment density correlation function of cluster orientation with matter in real space. In the top panel the blue and red points respectively show the correlations parallel and perpendicular to the major axes of the clusters $\xi_{mc}(r,\theta)$, while the black points shows the angularly averaged, conventional correlation function, $\xi_{mc}(r)$. The dotted curves are the LA model predictions. Note that the vertical axis mixes logarithmic and linear scalings. The inset provides the zoomed view on BAO scales. Here we also show how the NLA model (dashed curves) improves the fit around BAO scales compared to the LA model (dotted curves). The bottom panels show the the ratio, $\xi_{mc}(r,\theta)/\xi_{mc}(r)-1$. Essentially, the difference between LA (dotted) and NLA (dashed) models is the accuracy near BAO scales, $70<r<110\himpc$.
}
\label{fig:xidd_m}
\end{figure}

In the top panel of Fig.~\ref{fig:xidd_m}, we show the cluster-matter density cross-correlation function binned in $\theta$. Again, the inset zooms the correlation around the BAO scales \cite{Eisenstein:2005,Okumura:2008}. As found in Ref.~\cite{Faltenbacher:2012}, the BAO features are more enhanced in the correlation function perpendicular to the major axes of halos. The bottom panel shows the ratio, $\xi_{mc}(r,\theta)/\xi_{mc}(r)-1$, and thus the effect of IA is seen as the deviation from zero. The sign changes at  $\sim120\himpc$ because the conventional correlation function crosses zero there (see the upper panel). 

\begin{figure*}[bt]
\includegraphics[width=0.99\textwidth,angle=0,clip]{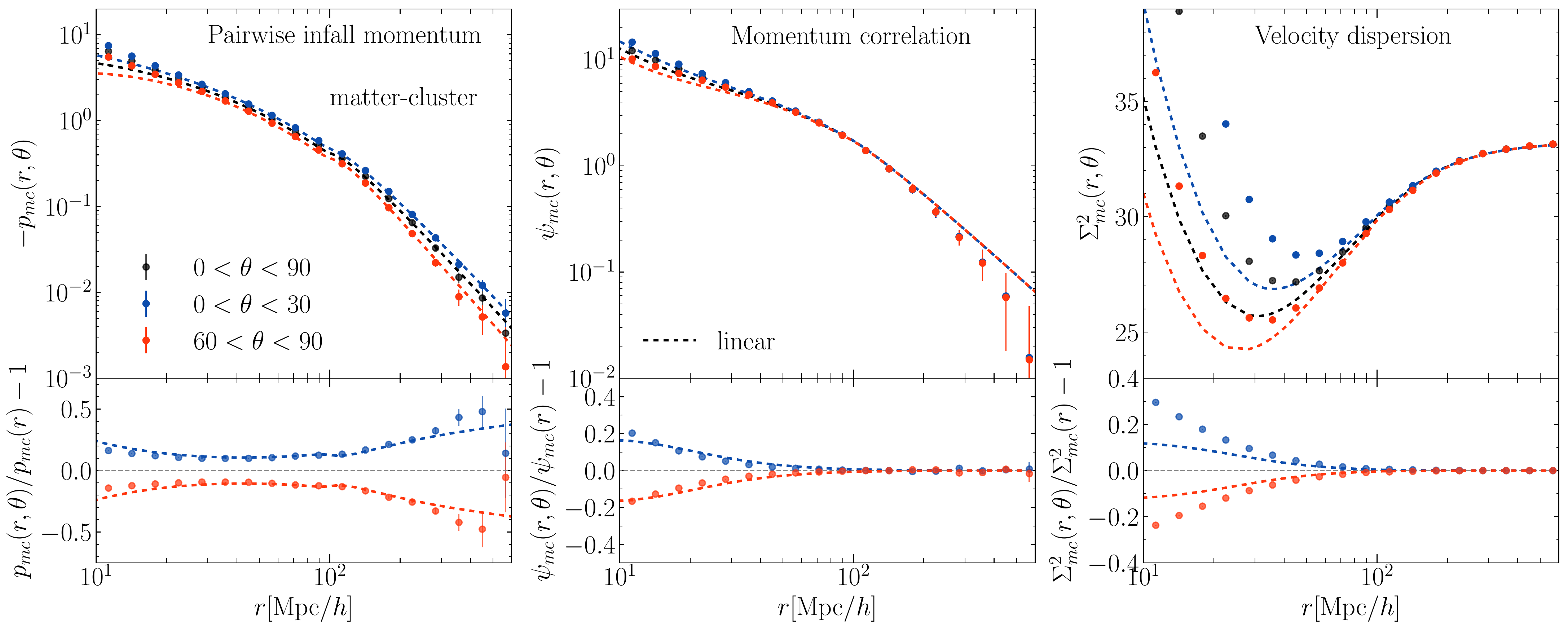}
\caption{Upper panels: Alignment pairwise mean infall momentum (left), alignment momentum correlation function (middle) and density-weighted alignment velocity dispersion (right), which respectively have units of $\himpc$, $\hsimpcs$ and $\hsimpcs$. Since the pairwise infall momentum has negative values at all scales, we show $-p_{mc}$ in the left panel. For the velocity dispersion we shift the linear theory prediction by a constant value vertically to match the measured dispersions in the large-scale limit (see text). The dotted curves are our model predictions for these velocity statistics based on the LA model.
Lower panel: Ratios of alignment velocity statistics to the corresponding conventional one, $p_{mc}(r,\theta)/p_{mc}(r)-1$ (left), $\psi_{mc}(r,\theta)/\psi_{mc}(r)-1$ (middle) and $\Sigma_{mc}^2(r,\theta)/\Sigma_{mc}^2(r)-1$ (right). Just like in the upper panels, the dotted curves show the LA model predictions.
}
\label{fig:xidt_tt_vv_m}
\end{figure*}

The next quantity to consider is the pairwise mean infall momentum. The upper-left panel of Fig.~\ref{fig:xidt_tt_vv_m} shows the alignment pairwise mean momentum $p_{mc}(r,\theta)$ and the lower-left panel shows its ratio with the conventional pairwise mean momentum, $p_{mc}(r,\theta)/p_{mc}(r)-1$. The sign of the statistic is negative over all the scales probed because objects approach each other on average through gravity. One can see the clear alignment signal up to very large scales, which indicates that dark matter residing along the major axis of a cluster tends to move faster toward the cluster than that perpendicular to the major axis. Note that the alignment signal in $p_{mc}$ persists beyond $r\sim 200\himpc$, up to $\sim 500\himpc$.  

We turn to another velocity statistic, the alignment momentum correlation function. Its numerical results are shown in the upper- and lower-middle panels of Fig.~\ref{fig:xidt_tt_vv_m}. The IA signal in the momentum correlation vanishes at $r\sim 100\himpc$, on scales much smaller than that of the density correlation. This is expected because the contributions of IAs to the momentum correlation function appear as products of two correlation functions, as seen in Eq.~(\ref{eq:acf_momentum_formula}). Finally, we show the density-weighted alignment velocity dispersion in the upper-right panel of Fig.~\ref{fig:xidt_tt_vv_m} and its ratio with the conventional one in the lower-right panel. As is the case with the alignment momentum correlation function, the effect of IA becomes negligible at large scales, as shown in the lower-right panel of Fig.~\ref{fig:xidt_tt_vv_m}.


\section{Comparison of model predictions to $N$-body results}\label{sec:comparison}
In this section, we compare the predictions of the LA model for the density and velocity alignment statistics to the measurements from $N$-body simulations obtained in Sec.~\ref{sec:measurements}. We first show the results for the density statistics, namely the GI and alignment density correlation functions in Sec.~\ref{sec:comparison_density}, and then those for the velocity statistics in Sec.~\ref{sec:comparison_velocity}. Since multiple terms contribute to the alignment signal in each velocity statistic, we discuss the individual contributions of these terms in Sec.~\ref{sec:each_term}. Section \ref{sec:biased_tracers} presents the results when biased objects are used rather than dark matter particles as a tracer of IA.

\subsection{Density statistics}\label{sec:comparison_density}

In Fig.~\ref{fig:gi_m_r_2gpc_z014}, the prediction of the LA model for the GI correlation is compared to the $N$-body result. The amplitude, $C_1$ in Eq. (\ref{eq:gi_la}), is determined by comparing the model to the GI measurement and minimizing the $\chi^2$ statistic. We find the best-fitting value to be $C_1\bar{\rho}/\bar{D}=1.50$. Note that the best-fitting value determined here is used for all the statistics presented below. The red dotted curve shows the prediction from the LA model. The result is consistent with the work of Ref.~\cite{Blazek:2011}, and we extend the modeling to larger scales. We find perfect agreement between the measurement and the LA model at $20<r<640\himpc$. As shown in the inset of Fig.~\ref{fig:gi_m_r_2gpc_z014}, there is a small but non-negligible discrepancy between the measurement and the LA model prediction around BAO scales. However, the NLA model with the PT-based nonlinear correction to the spectra using the \regpt code \cite{Taruya:2012}, depicted by the dashed curve, dramatically improves the agreement, consistent with the result of Ref.~\cite{Chisari:2013}.

The dotted curves in Fig.~\ref{fig:xidd_m} are the result of the LA model prediction for the alignment density correlation function. 
Note again that for the parameter $C_1$, we use the same value as determined from the GI correlation. 
Since the alignment correlation is equivalent to the GI correlation, 
the accuracy of the LA model is the same as that for the GI correlation. 
And thus, as shown in the inset of the upper panel, the LA model fails to reproduce the nonlinear smearing effect around the BAO peak while the NLA model significantly improves the accuracy. 
The lower panel of Fig.~\ref{fig:xidd_m} shows that although the LA model overpredicts the amplitude of IA at $r\sim 80 \himpc$, the NLA model perfectly predicts it. 

Our analysis reproduces the modeling result of Ref.~\cite{Blazek:2011} but we extend it to larger scales, $>100\himpc$. Since the alignment correlation along the major axis ($\theta \sim 0^\circ$) at such scales is close to zero, the ratio $\xi_{mc}(r,\theta)/\xi_{mc}(r)-1$ becomes quite noisy as seen in the bottom panel. 
On the other hand, the correlation perpendicular to the major axis ($\theta \sim 90^\circ$) can be accurately modeled by the LA model up to $\sim 400\himpc$.

\subsection{Velocity statistics}\label{sec:comparison_velocity}
We now discuss the predictions of the LA model for the velocity alignment statistics we proposed. 
While the measured VI correlation function is density weighted, our model prediction in Eq.~(\ref{eq:vi_la}) is volume weighted, $\xi_{v_m+}$.
Thus, we also compare our model to the measurement of $\hat{\xi}_{v_m+}\equiv \xi_{p_m+}(1+\xi_{\delta_m\delta_c})^{-1}$.
We do not plot the measurement of $\hat{\xi}_{v_m+}$ because not only they are not the exactsame quantities (see Appendix B of Ref.~\cite{Okumura:2014}) but there is also a negligible difference between 
the measurements of $\xi_{p_m+}$ and $\hat{\xi}_{v_m+}$ except at $r<25\himpc$. 
The LA model for the VI correlation with the same value of $C_1$ as the GI correlation predicts the measurement on large scales.
The model, however, fails to predict the VI correlation at $r<50\himpc$, which is a relatively larger scale than the GI correlation. 

\begin{figure*}[bt]
\includegraphics[width=0.99\textwidth,angle=0,clip]{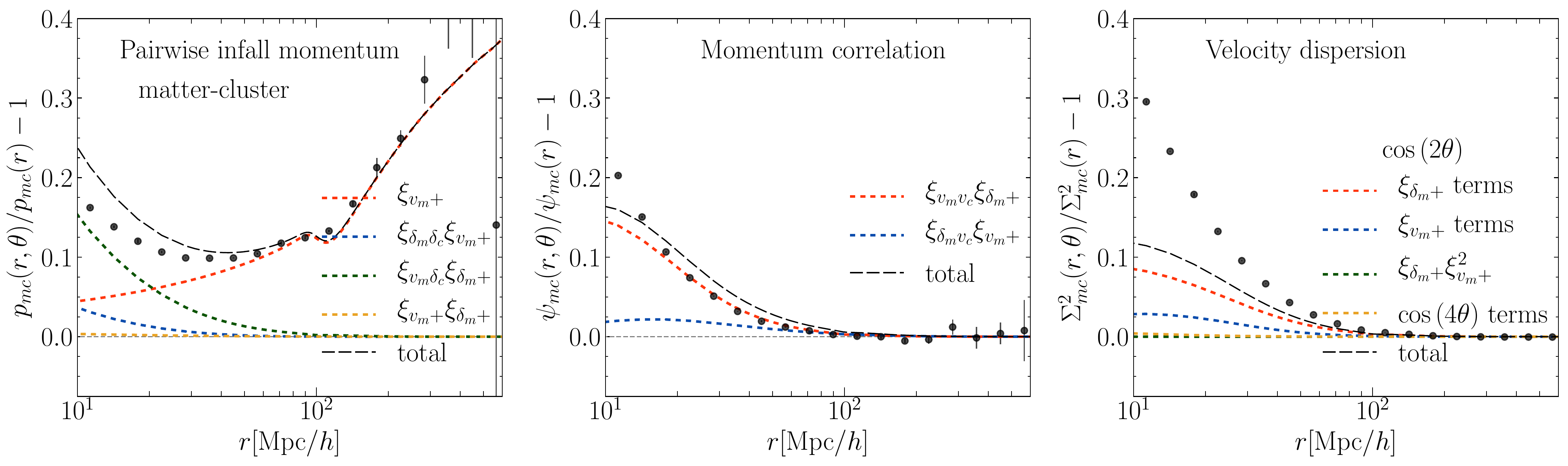}
\caption{
Contributions of each term to the IA in the pairwise infall momentum (left), momentum correlation (middle) and density-weighted velocity dispersion (right) for $0<r<30^\circ$.
In each panel, the points are the $N$-body measurements, which are the same data points as the blue points shown in the lower panels of Fig.~\ref{fig:xidt_tt_vv_m}. 
In the left panel, contributions of the terms with $\xi_{v_m+}$, $\xi_{\delta_m\delta_c}\xi_{v_m+}$, $\xi_{\delta_m\delta_c}\xi_{\delta_m+}$, and $\xi_{v_m+}\xi_{\delta_m+}$ in Eq. (\ref{eq:acf_infall_formula})
are shown as red, blue, green and yellow dotted curves, respectively, and the sum of these terms is shown as black dashed curves. 
In the middle panel, two terms contribute to the IA of the momentum correlation, and the $\xi_{v_mv_c}\xi_{\delta_m+}$ and $\xi_{\delta_mv_c}\xi_{v_m+}$ terms are shown by the red and blue curves, respectively. 
Because there are many terms contributing to the IA of the density-weighted velocity dispersion, in the right panel we show the contributions of the $\cos{(2\theta)}$ terms proportional to $\xi_{\delta_m+}$ (red), $\xi_{v_m+}$ (blue) and $\xi_{\delta_m+}\xi_{v_m+}^2$ (green), and the sum of the $\cos{(4\theta)}$ terms (yellow). Since the size of the $\theta$ bin is $30^\circ$, there is no contribution from the $\cos{(6\theta)}$ term.
}
\label{fig:vel_ia_m_z014}
\end{figure*}

The LA model prediction for the alignment pairwise mean momentum is shown as the dotted curves and compared to the result from $N$-body simulations in the upper- and lower-left panels in Fig.~\ref{fig:xidt_tt_vv_m}. The conventional pairwise mean momentum is consistent with linear theory, $[\xi_{\delta_mv_c}(r)-\xi_{v_m\delta_c}(r)]$, on large scales and starts to deviate at the scale $r\sim 30\himpc$, as studied in Refs.~\cite{Bhattacharya:2008,Okumura:2014}. The LA model can accurately predict the IA of the pairwise mean momentum up to similar scales. As shown in the lower-left panel of Fig.~\ref{fig:xidt_tt_vv_m}, the large-scale velocity alignment in $N$-body simulations perfectly matches the model prediction and the nonzero alignment signal is detected up to $\sim 500\himpc$. Compared to the result in Fig.~\ref{fig:xidd_m}, this implies that the alignment signal can be better probed by using the phase-space information. While the LA model fails to predict the nonlinearity of the pairwise infall momentum on small scales, it is canceled out by taking the ratio, $p_{mc}(r,\theta)/p_{mc}(r)-1$, and the alignment effect itself can be well captured by the LA model on such scales, as demonstrated in the lower-left panel of Fig.~\ref{fig:xidt_tt_vv_m}. Many terms contribute to the IAs of the pairwise infall momentum [Eq.~(\ref{eq:acf_infall_formula})], and the contribution of each term will be discussed in Sec.~\ref{sec:each_term}.

Let us now present our prediction for the alignment momentum correlation function based on the LA model. First, we study the prediction for the conventional momentum correlation, 
$\psi_{AB}(r)= (1+\xi_{\delta_A\delta_B})\xi_{v_A v_B }+\xi_{\delta_A v_B}\xi_{v_A\delta_B}$,
shown as the black dotted curve and compared to the $N$-body result in the middle set of Fig.~\ref{fig:xidt_tt_vv_m}. The prediction for the alignment momentum correlation function is depicted by the blue ($0<\theta<30^\circ$) and red ($60<\theta<90^\circ$) curves. As seen in the lower panel, there is no large-scale IA effect in the momentum correlation, consistent with the measurement. The measured momentum correlation has a lower amplitude than the linear prediction on very large scales, particularly at $r>300\himpc$. This is caused by the finite simulation box (survey) size, which is more significant for the velocity field than the density field (see, e.g., Refs.~\cite{Okumura:2014,Sugiyama:2016}). Investigating the finite-volume effect is beyond the scope of this paper and it will be discussed in further detail in our future work. There are several terms which contribute to the IA signal seen on small scales [Eq.~(\ref{eq:acf_momentum_formula})], and they will be addressed in Sec.~\ref{sec:each_term} below. 

Finally, the LA model is tested for the density-weighted alignment velocity dispersion in the upper- and lower-right panels of Fig.~\ref{fig:xidt_tt_vv_m}. It is well-known that the linear theory prediction for the one-dimensional (1D) velocity dispersion, $\sigma_{v_A}=3.49\himpc$ (at $z=0.306)$, causes a constant offset for the density-weighted velocity dispersion \cite{Reid:2011,Vlah:2012,Okumura:2014}. Thus, an extra term needs to be added to $\sigma_{v_A}^2$, $\sigma_{v_A}^2 \to \sigma_{v_A,{\rm nl}}^2=\sigma_{v_A}^2+\Delta \sigma_{v_A}^2$. Since modeling the nonlinear velocity dispersion is not within the scope of this paper, to match the amplitude of the measured velocity dispersion on large scales we simply add a constant so that $\sigma_{v,{\rm nl}} \equiv \sqrt{(\sigma_{v_m,{\rm nl}}^2+\sigma_{v_c,{\rm nl}}^2)/2}=4.08 \himpc$. Just for comparison, the 1D velocity dispersions directly computed from the simulations are $\sigma_{v_m,{\rm nl}}=4.17\himpc$ and $\sigma_{v_c,{\rm nl}}=3.95\himpc$, and thus $\sigma_{v,{\rm nl}} = 4.06 \himpc$. Although the two $\sigma_{v,{\rm nl}}{}'s$ are not the exact same quantities, adding the constant to have $\sigma_{v,{\rm nl}}$ similar to the directly measured value would be a reasonable manipulation. 
As seen in the upper panel, the  behavior of the alignment velocity dispersion is qualitatively captured by the LA model. However, the model significantly underpredicts the measurement as in the lower panel. We discuss the individual contributions in Sec.~\ref{sec:each_term}. 

\subsection{Contribution of each term} \label{sec:each_term}
We look into the modeling of the velocity alignment statistics in more detail. From the left to right panels of Fig.~\ref{fig:vel_ia_m_z014}, the points with error bars show the measured ratios, $p_{mc}(r,\theta)/p_{mc}(r)-1$, $\psi_{mc}(r,\theta)/\psi_{mc}(r)-1$ and $\Sigma^2_{mc}(r,\theta)/\Sigma^2_{mc}(r)-1$ with $0<\theta<30^\circ$. They are respectively the same as the blue points in the lower panels of Fig.~\ref{fig:xidt_tt_vv_m} from left to right.

The pairwise momentum has a linear-order contribution of IA, proportional to the VI correlation, $\xi_{v_m+}$, depicted by the red curves in the upper panel. This term dominates the signal at scales larger than $r\sim 50\himpc$. On small scales, on the other hand, many terms contribute. The major contribution is the term proportional to $\xi_{\delta_m\delta_c}\xi_{v_m+}$, followed by the term of $\xi_{\delta_m\delta_c}\xi_{\delta_m+}$, shown by the blue and green curves, respectively. The $\cos{(4\theta)}$ moment, proportional to $\xi_{v_m+}\xi_{\delta_m+}$, has a negligible contribution (yellow curve).

\begin{figure}[b]
\includegraphics[width=0.49\textwidth,angle=0,clip]{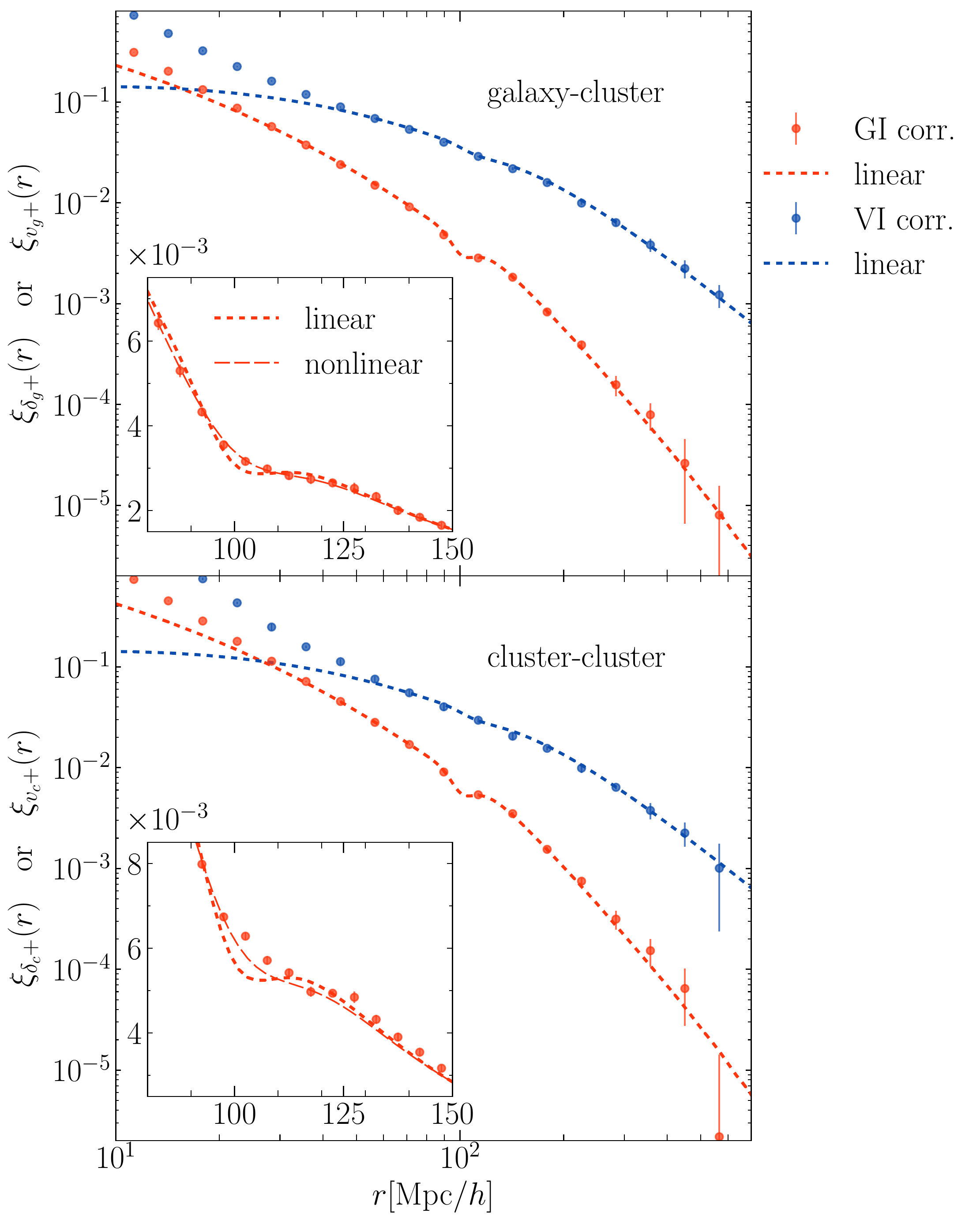}
\caption{Same as in Fig.~\ref{fig:gi_m_r_2gpc_z014}, but for the GI and VI correlation functions for cluster shapes whose density/velocity fields are sampled by galaxies (top) and clusters (bottom).
}
\label{fig:gi_h_r_2gpc_z014}
\end{figure}

\begin{figure}[b]
\includegraphics[width=0.38\textwidth,angle=0,clip]{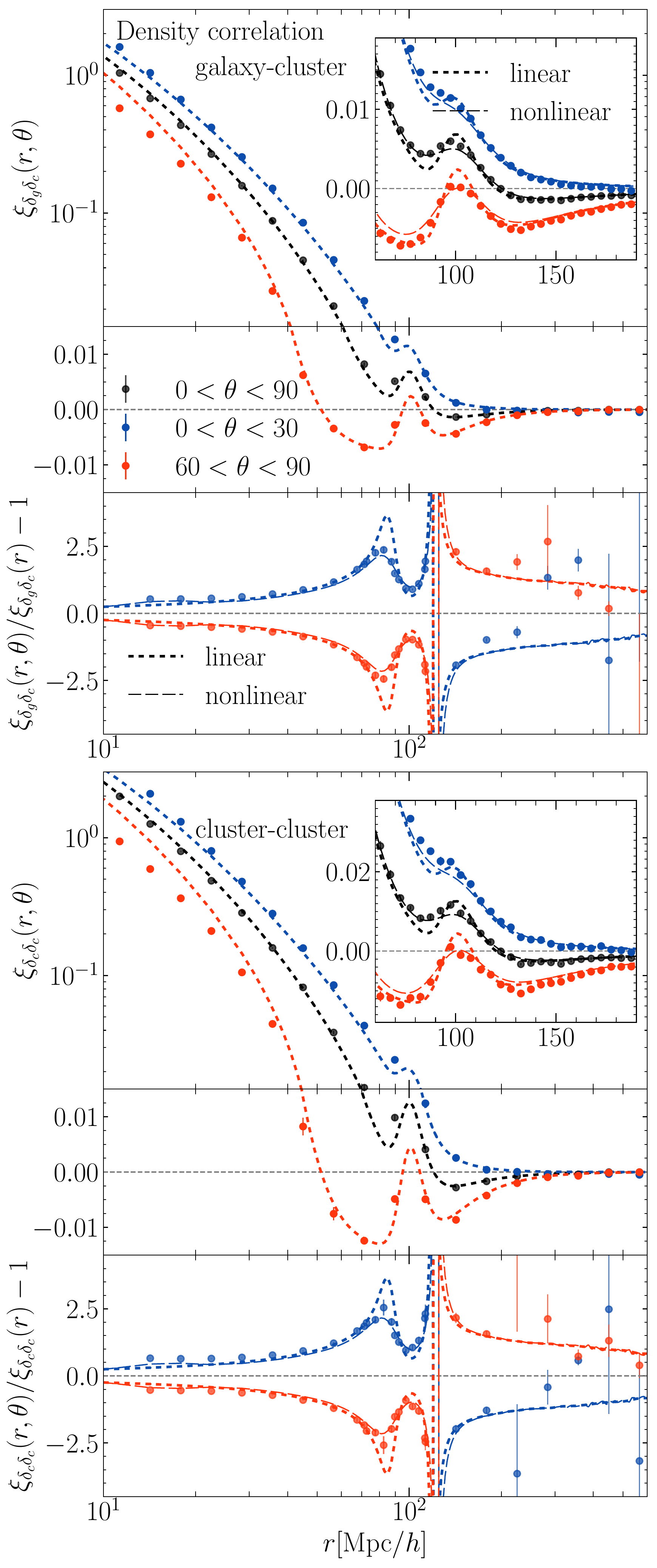}
\caption{
Same as in \ref{fig:xidd_m}, but for the alignment density correlation function of cluster orientation with galaxies (upper set) and clusters (lower set). The top panel of each set shows the correlation function itself, $\xi_{Ac}(r,\theta)$  where $A=\{g,c\}$, while the bottom panel shows the ratios, $\xi_{Ac}(r,\theta)/\xi_{Ac}(r)-1$.  
}
\label{fig:xidd_h}
\end{figure}

\begin{figure*}[bt]
\includegraphics[width=0.99\textwidth,angle=0,clip]{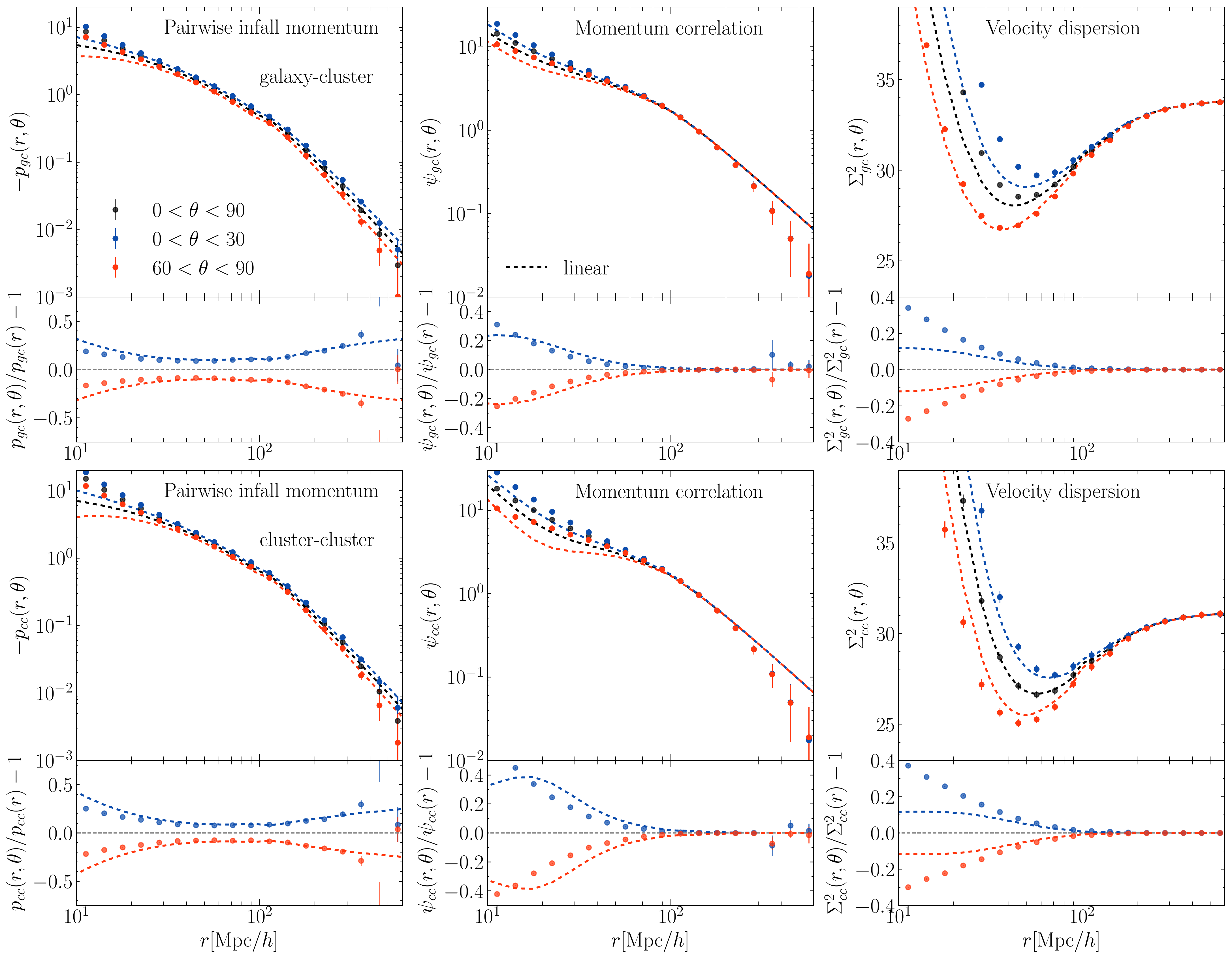}
\caption{Same as in Fig.~\ref{fig:xidt_tt_vv_m}, but in the upper and lower sets, galaxies and clusters are respectively used instead of dark matter as a tracer of the velocity field for the velocity alignment statistics.
}
\label{fig:xidt_tt_vv_h}
\end{figure*}

For the momentum correlation function, there are two terms which contribute to the IA effect, $\xi_{v_Av_B}(\vecr)\xi_{\delta_A +}(\vecr)$ and $\xi_{\delta_A v_B }(\vecr)\xi_{v_A  +}(\vecr)$, shown as the red and blue curves in the middle panel of Fig.~\ref{fig:vel_ia_m_z014}. As expected, all the terms are higher order and vanish on large scales $r\sim 100\himpc$. Many terms contribute to the IA in the density-weighted velocity dispersion. In the right panel of Fig.~\ref{fig:vel_ia_m_z014} we show contributions of the $\cos{(2\theta)}$ terms proportional to $\xi_{\delta_m+}$ (red), $\xi_{v_m+}$ (blue) and $\xi_{\delta_m+}\xi_{v_m+}^2$ (green), and the sum of the $\cos{(4\theta)}$ terms (yellow). Note that since our statistics are binned by $30^\circ$ in $\theta$, the term proportional to $\cos{(6\theta)}$ vanishes. Just as in the case of the momentum correlation, the IA effect in the velocity dispersion becomes zero at large scales. However, unlike the former velocity statistics, the LA model significantly underpredicts the IA effect on small scales. This implies that the nonlinearity of IA needs to be taken into account properly for the precise modeling of the alignment velocity dispersion on small scales.

\begin{figure*}[bt]
\includegraphics[width=0.99\textwidth,angle=0,clip]{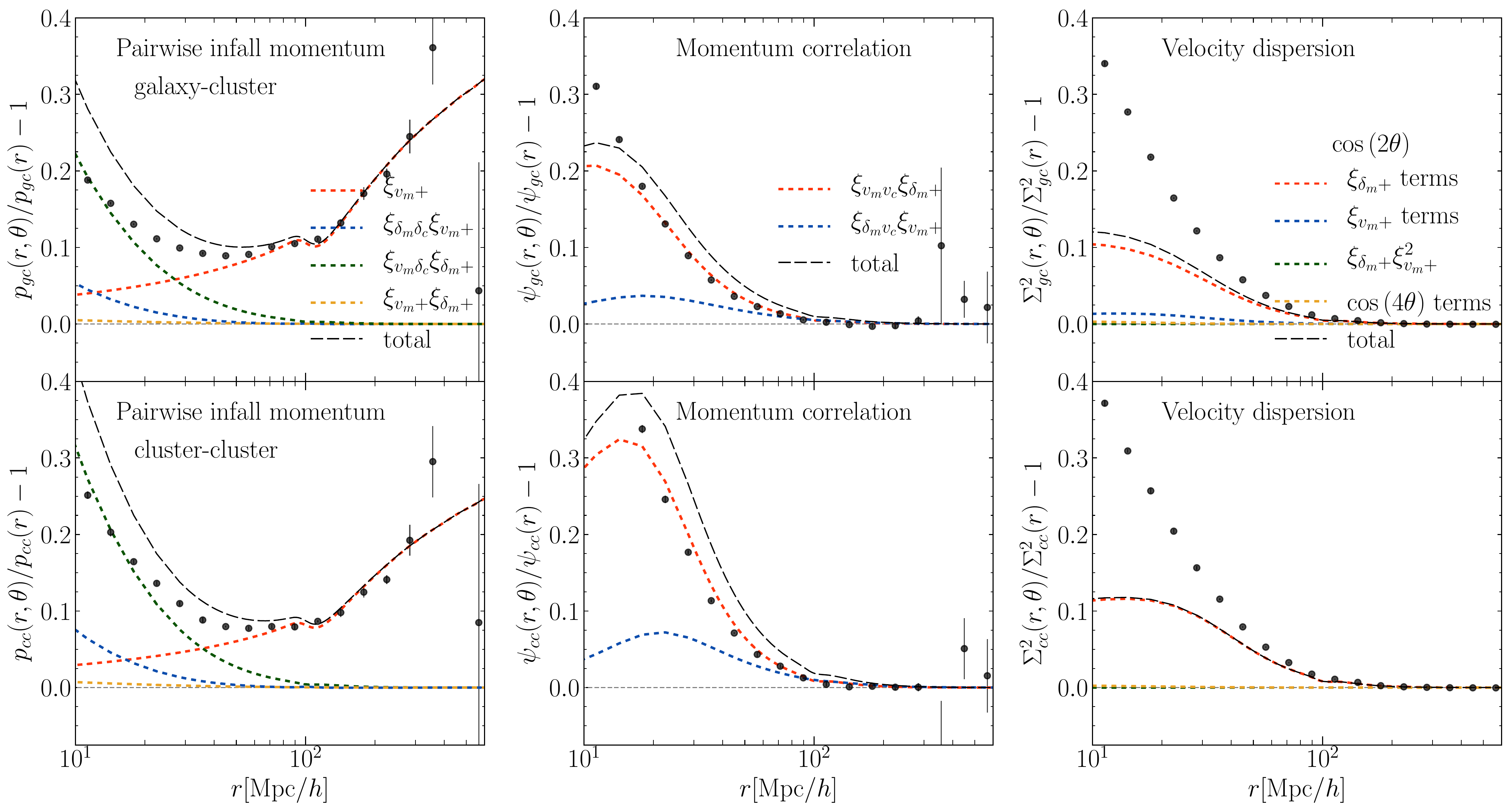}
\caption{
Same as in Fig.~\ref{fig:vel_ia_m_z014}, but in the upper and lower sets, galaxies and clusters are respectively used instead of dark matter as a tracer of the velocity field for the velocity alignment statistics.
}
\label{fig:vel_ia_h_z014}
\end{figure*}

\subsection{Biased tracers}\label{sec:biased_tracers}
So far we have studied the IAs of halos based on the density and velocity fields of dark matter particles relative to the cluster orientations. In this subsection we investigate this effect using the biased objects as a tracer of IA, namely the cross correlations between the cluster orientations with density/velocity fields of clusters or galaxies. Even for the alignment statistics of these biased objects, we use the  single number $C_1$ determined for the matter-cluster GI correlation function in Sec.~\ref{sec:comparison_density}. The upper and lower panels of Figs.~\ref{fig:gi_h_r_2gpc_z014} -- \ref{fig:vel_ia_h_z014} show respectively the results for galaxy-cluster and cluster-cluster statistics, corresponding to those for matter-cluster statistics in Figs.~\ref{fig:gi_m_r_2gpc_z014} -- \ref{fig:vel_ia_m_z014}.

The upper panel of Fig.~\ref{fig:gi_h_r_2gpc_z014}, shows the GI correlation between galaxy density and cluster orientation. On large scales where linear theory holds, the GI cross correlation is related to that with dark matter by a linear relation, $\xi_{\delta_g+}=b_g \xi_{\delta_m+}$ \cite{Hirata:2007}, where the galaxy bias $b_g$ is determined to be $b_g\sim 1.70$ (Sec.~\ref{sec:nbody}). Just like in Fig.~\ref{fig:gi_m_r_2gpc_z014}, the LA model predicts the $N$-body result of the GI correlation, and the NLA model with the nonlinear power spectrum based on PT improves the accuracy around BAO scales as shown in the inset of the upper panel of Fig.~\ref{fig:gi_h_r_2gpc_z014}. However, if clusters are used as a tracer of the IA, the nonlinear smearing of the measured BAO signal still slightly deviates from the the NLA model prediction as shown in the lower panel of Fig.~\ref{fig:gi_h_r_2gpc_z014}. This could be due to the nonlinearity of the bias. We show the alignment galaxy-cluster and cluster-cluster density correlation function in Fig.~\ref{fig:xidd_h}. This signal has been measured both in simulations and observations \cite{Faltenbacher:2009,Schneider:2012,Li:2013}, and the LA model fitting was performed in Ref. \cite{Blazek:2011}. Again, our analysis extends the previous results toward larger scales. The NLA model improves the agreement with the measured alignment correlation around BAO scales. However, the BAO features are slightly more significant than the NLA model prediction with the linear bias ansatz, unlike the matter-cluster correlation. It is known that in peak theory the BAO peak is amplified for high peaks which correspond to highly biased objects such as luminous galaxies and clusters \cite{Desjacques:2008,Desjacques:2010}. Thus, the result obtained here is consistent with the peak theory prediction. 

We show the VI correlation function, namely the correlation of the cluster orientation with galaxy and cluster velocities, respectively, in the upper and lower panels of Fig.~\ref{fig:gi_h_r_2gpc_z014}. Because the VI correlation does not depend on the bias in the LA model, the blue dashed and dotted curves are equivalent to those in Fig.~\ref{fig:gi_m_r_2gpc_z014}. Perfect agreement between the measurements and the corresponding model predictions is achieved at $r>50\himpc$ up to the largest scale probed in this paper, $r\sim 600\himpc$. 

The alignment pairwise infall momenta for biased tracers are shown in the left panels of Fig.~\ref{fig:xidt_tt_vv_h}. By comparing them to the matter-cluster alignment pairwise momentum in the left panel of Fig.~\ref{fig:xidt_tt_vv_m}, one can see that the IA of the pairwise infall momentum slightly depend on the bias of the density/velocity tracer, and more biased objects have smaller alignments. The middle panel of Fig.~\ref{fig:xidt_tt_vv_h} presents the alignment momentum correlation function for these biased objects. Just like the alignment matter-cluster momentum correlation, the effect of IA vanishes at scales beyond $r\sim 100\himpc$. On smaller scales the deviation of the shape of the measured momentum correlation from the linear theory becomes more significant for more biased tracers, although the IA of the galaxy-cluster momentum correlation is still consistent with the LA model prediction even at such small scales, as shown in the bottom of the upper-middle panel of Fig.~\ref{fig:xidt_tt_vv_h}. The result for the alignment density-weighted pairwise velocity dispersion of the biased objects is shown in the upper- and lower-right panels of Fig.~\ref{fig:xidt_tt_vv_h}. Once again, the nonlinear correction to the linear-theory velocity dispersion, $\Delta \sigma_{v_A}^2 = \sigma_{v_A,{\rm nl}}^2-\sigma_{v_A}^2$, is added by hand to match the measured simulation result in the large-scale limit. Since the 1D velocity dispersion of clusters directly measured from the simulations, $\sigma_{v_c,{\rm nl}}=3.95\himpc$, provides a perfect match with the large-scale amplitude of $\Sigma_{cc}^2$, we simply use the value of $\Delta\sigma_{v_c}$ which gives this value of $\sigma_{v_c,{\rm nl}}$. For $\Sigma_{gc}^2$, we choose $\sigma_{v,{\rm nl}} \equiv \sqrt{(\sigma_{v_g,{\rm nl}}^2+\sigma_{v_c,{\rm nl}}^2)/2}=4.12\himpc$, which is slightly higher than the directly computed value, $\sigma_{v,{\rm nl}}=4.06\himpc$. 

\begin{figure*}[bt]
\includegraphics[width=0.99\textwidth,angle=0,clip]{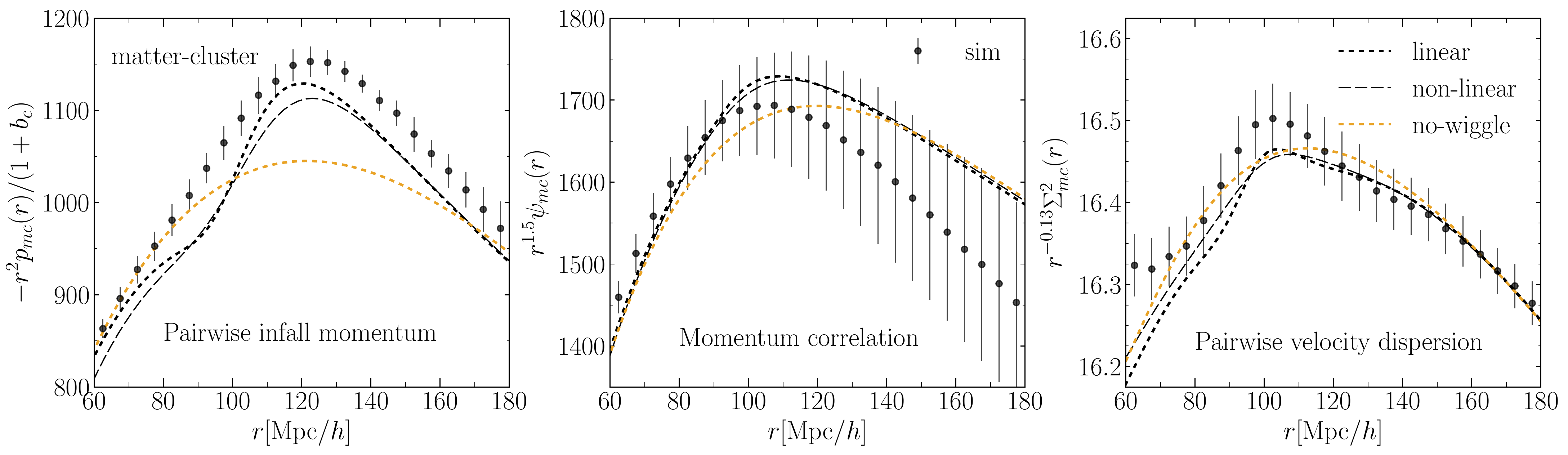}
\caption{
BAO features in velocity statistics shown in the upper panels of Fig.~\ref{fig:xidt_tt_vv_m}. We plot the pairwise infall momentum (left), momentum correlation (middle) and pairwise velocity dispersion (right) between matter and clusters, multiplied by some powers of $r$ to make the slopes flatter. The black points and dotted curves are respectively the $N$-body measurements and the model predictions based on linear theory, which are the same as those in the upper panels of Fig.~\ref{fig:xidt_tt_vv_m}. The black dashed curves are the PT-based nonlinear model, while the yellow dotted curves are the linear predictions with the BAO wiggles smeared out. 
}
\label{fig:vel_nia_bao_m_z014}
\end{figure*}

Figure \ref{fig:vel_ia_h_z014} is similar to Fig.~\ref{fig:vel_ia_m_z014}, but it presents the ratios of the IA velocity statistics to the corresponding conventional ones, $X_{gc}(r,\theta)/X_{gc}(r)-1$ (upper row) and $X_{cc}(r,\theta)/X_{cc}(r)-1$ (lower row), where $X=\{p,\psi,\Sigma^2\}$ from left to right. The points with the error bars and the black curves are the same as the blue points and curves in the bottom parts of the upper panels in Fig.~\ref{fig:xidt_tt_vv_h}. The other curves show the contribution of each term to the total IA. As shown in the left panels, the large-scale IA signal in the pairwise infall momentum is well predicted by the LA model even for biased tracers. The alignment signals in the momentum correlation function at scales $r<100\himpc$ get stronger for tracers with larger biases, as demonstrated in the middle panels of Fig.~\ref{fig:vel_ia_h_z014} compared to the middle panel of Fig.~\ref{fig:vel_ia_m_z014}. Although our prediction uses the LA model with the linear power spectrum, such a trend can be well captured. The behavior of the IA effect measured for the density-weighted velocity dispersion is not very different for different tracers, even from dark matter, as in the right panels of Figs.~\ref{fig:vel_ia_m_z014} and \ref{fig:vel_ia_h_z014}. Thus, as in the dark matter case, our LA model largely underpredicts the measurement at scales less than $r \simeq 60\himpc$.

\section{Baryon acoustic oscillation features in velocities}\label{sec:bao}
In the previous section we studied the BAO features encoded in only the density and its alignment statistics. However, it is natural to expect such features in the velocity statistics as well. The BAO features encoded in the velocity statistics in Fourier space have already been discussed in Ref.~\cite{Okumura:2014}. In Sec.~\ref{sec:bao_nia} we present such BAO features for the corresponding configuration space. Then in Sec.~\ref{sec:bao_ia} we demonstrate that the VI correlation function and alignment pairwise infall momentum also contain the BAO information. 

\subsection{BAOs in conventional velocity statistics}\label{sec:bao_nia}
Figure \ref{fig:vel_nia_bao_m_z014} shows the conventional pairwise infall momentum, momentum correlation function, and density-weighted velocity dispersion of clusters with matter from left to right. The black points are the same as those in the upper panels of Fig.~\ref{fig:xidt_tt_vv_m}, but these quantities are multiplied by some powers of the separation, $r^n$, in order to flatten the results near the BAO scales. The black dotted and dashed curves are the linear and nonlinear model predictions, the latter of which is computed computed using the \regpt code. The yellow curve is computed by the linear model with the linear power spectrum without BAO wiggles, $P_{\delta_m\delta_m}^{\rm nw}$, where the superscript ``nw'' denotes a ``no-wiggle'' power \cite{Eisenstein:1998}.

For the pairwise infall momentum in the left panel of Fig.~\ref{fig:vel_nia_bao_m_z014}, one can see a small but systematic offset of the amplitude between the $N$-body result and any of the theoretical predictions. Since the correlation between different separation bins in the velocity field in configuration space is stronger than that in the density field, it is possible for the correlation to be systematically shifted vertically. Subtracting a small constant from the measured $p_{mc}$ at all scales indeed provides the agreement with the PT-based model visually perfect (see Fig.~2. of Ref.~\cite{Eisenstein:2005} for the same trend found in the redshift-space correlation function which also contains velocity information). However, our intent here is simply to demonstrate that the BAO information is indeed encoded in the velocity statistics, not to test which model is preferred. Thus, we will leave the more detailed modeling for future work. Because the results for the other two pairwise infall momenta, $p_{gc}(r)$ and $p_{cc}(r)$ are more or less equivalent to that for $p_{mc}(r)$, we do not show them here. 

\begin{figure}[bt]
\includegraphics[width=0.34\textwidth,angle=0,clip]{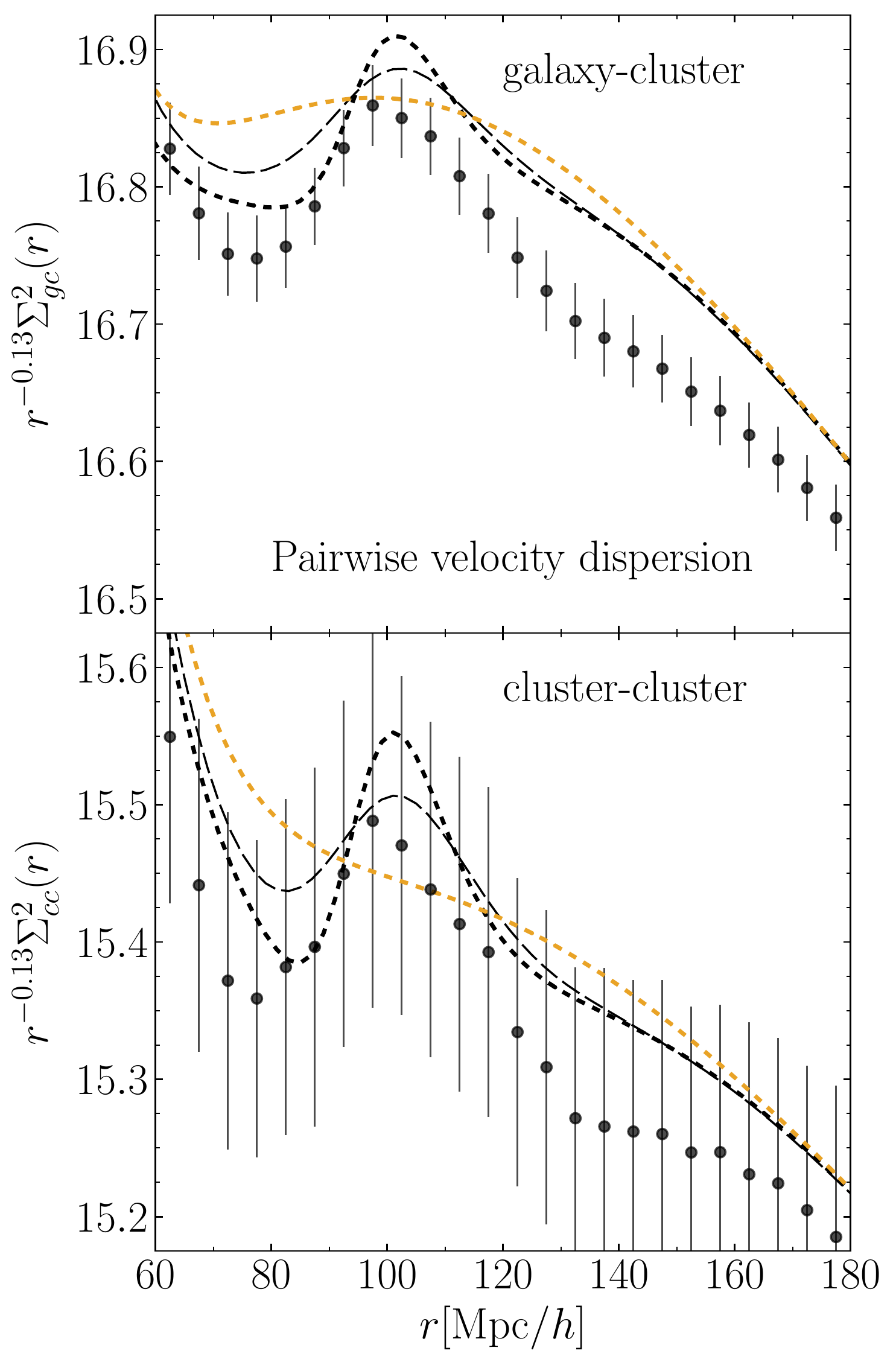}
\caption{
Same as in the right panel of Fig.~\ref{fig:vel_nia_bao_m_z014} but for the pairwise velocity dispersion for galaxy-cluster (top) and cluster-cluster (bottom) pairs. 
}
\label{fig:vel_nia_bao_h_z014}
\end{figure}

As shown in the middle panel of Fig.~\ref{fig:vel_nia_bao_m_z014}, the difference between the predictions with BAOs in linear and nonlinear perturbation theory is quite small. It is even not easy to distinguish between the models and the linear prediction without BAOs. Moreover, the difference is smaller than the finite-volume effect which becomes significant at scales $r\geq 100\himpc$, although the measurement is consistent with all three predictions within the error bars. Thus it will be of essential importance to fully understand the effect of the finite survey volume for the momentum correlation function if we want to use the BAO information in the correlation. 

As is the case with the momentum correlation function, it is hard to distinguish between different theoretical models for the pairwise velocity dispersion at the BAO scales as seen in the right panel of Fig.~\ref{fig:vel_nia_bao_m_z014}, although one can see a bump that could be due to the BAO signal. Note that, as described in Sec.~\ref{sec:comparison_velocity}, a constant, $\Delta\sigma_v^2$, is added to the prediction of the velocity dispersion to match with the measurement in the large-scale limit. For this statistics, it is interesting to see the case of highly biased objects because $\Sigma_{AB}^2(r)$ contains a term proportional to $\sigma_v^2\xi_{\delta_A\delta_B}(r)=\sigma_v^2b_Ab_B\xi_{\delta_m\delta_m}(r)$ [see Eq. (\ref{eq:velocity_disp_formula})]. We show the velocity dispersion for galaxy-cluster pairs, $\Sigma_{gc}^2$ in the upper panel of Fig.~\ref{fig:vel_nia_bao_h_z014}. Here the bump caused by BAOs is more prominent than that in the matter-cluster pairwise velocity dispersion as expected. There is a small ($< 1\%$) offset between the $N$-body result and our predictions because the constant $\Delta \sigma_v^2$ is added so that they match at larger scales, $r>200\himpc$. The BAO bump is more enhanced for the cluster-cluster velocity dispersion because of the $k^2$-dependent bias predicted by peak theory \cite{Desjacques:2008,Desjacques:2010}, as shown in the lower panel of Fig.~\ref{fig:vel_nia_bao_h_z014}. However, the measurement is so noisy due to the sparseness of clusters that it is hard to conclude anything concrete based on this result.

\subsection{BAOs in alignment velocity statistics}\label{sec:bao_ia}
\begin{figure}[t]
\includegraphics[width=0.325\textwidth,angle=0,clip]{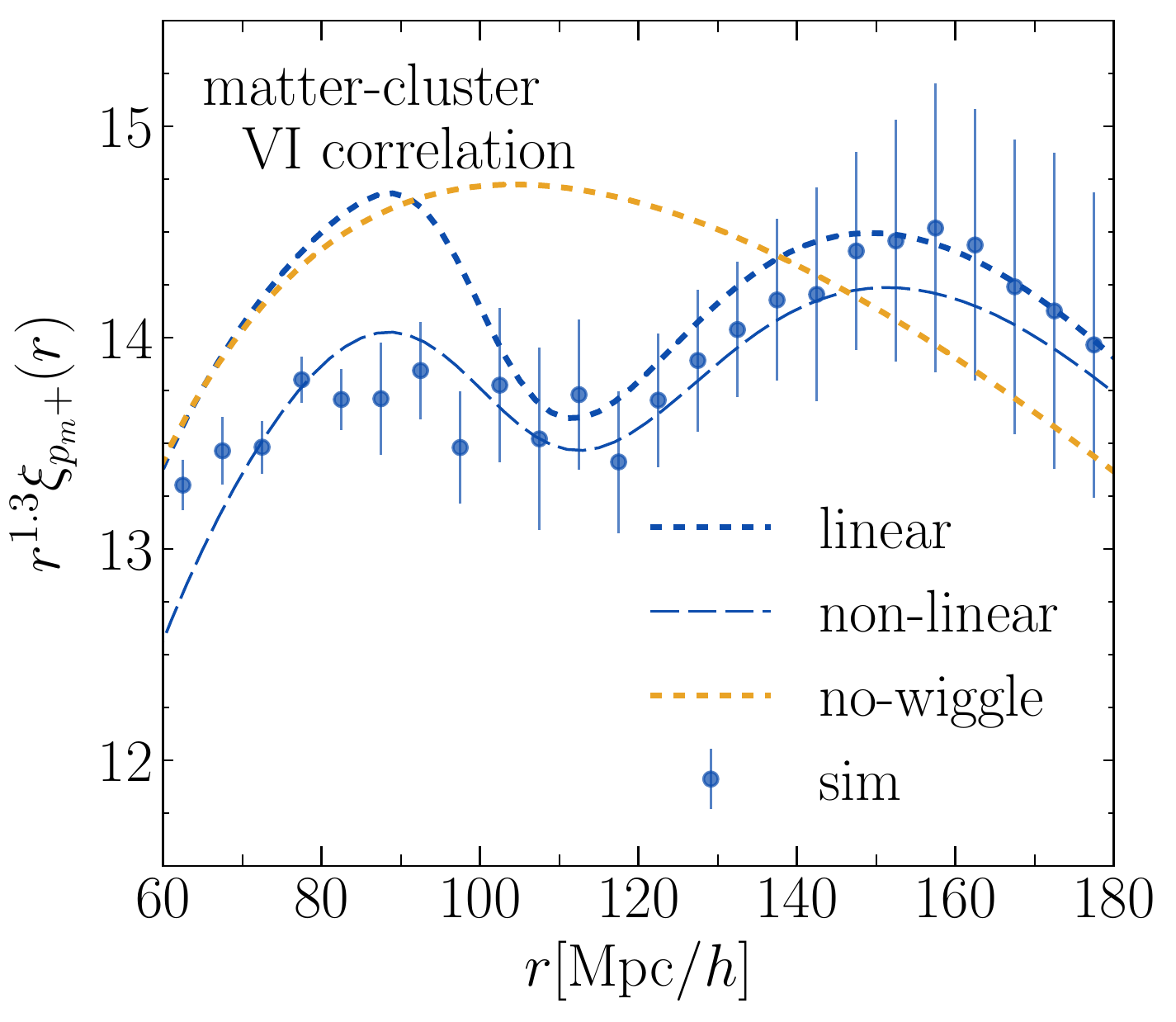}
\caption{
Same as in Fig.~\ref{fig:gi_m_r_2gpc_z014} but for the matter-cluster VI correlation function times $r^{1.3}$. The blue points and dotted curve are the same as those in Fig.~\ref{fig:gi_m_r_2gpc_z014}, i.e., the $N$-body result and the LA model prediction, respectively. The blue dashed and yellow dotted curves are the predictions with the NLA and no-wiggle LA models, respectively.
}
\label{fig:vel_ia_bao_m_vi_z014}
\end{figure}

Here, let us extend the analysis of BAOs in the former subsection to velocity statistics with IA. Figure \ref{fig:vel_ia_bao_m_vi_z014} plots the VI correlation function, the same as in Fig.~\ref{fig:gi_m_r_2gpc_z014}, but multiplied by $r^{1.3}$. Just like the GI correlation, BAOs contribute to the VI correlation negatively as emphasized in the figure. While the yellow dotted curve is the theoretical prediction of the LA model with BAOs smeared out, the blue dotted and dashed curves are respectively those of LA and NLA models with BAOs included. Obviously, our measurement prefers the models with BAOs. 

\begin{figure}[t]
\includegraphics[width=0.43\textwidth,angle=0,clip]{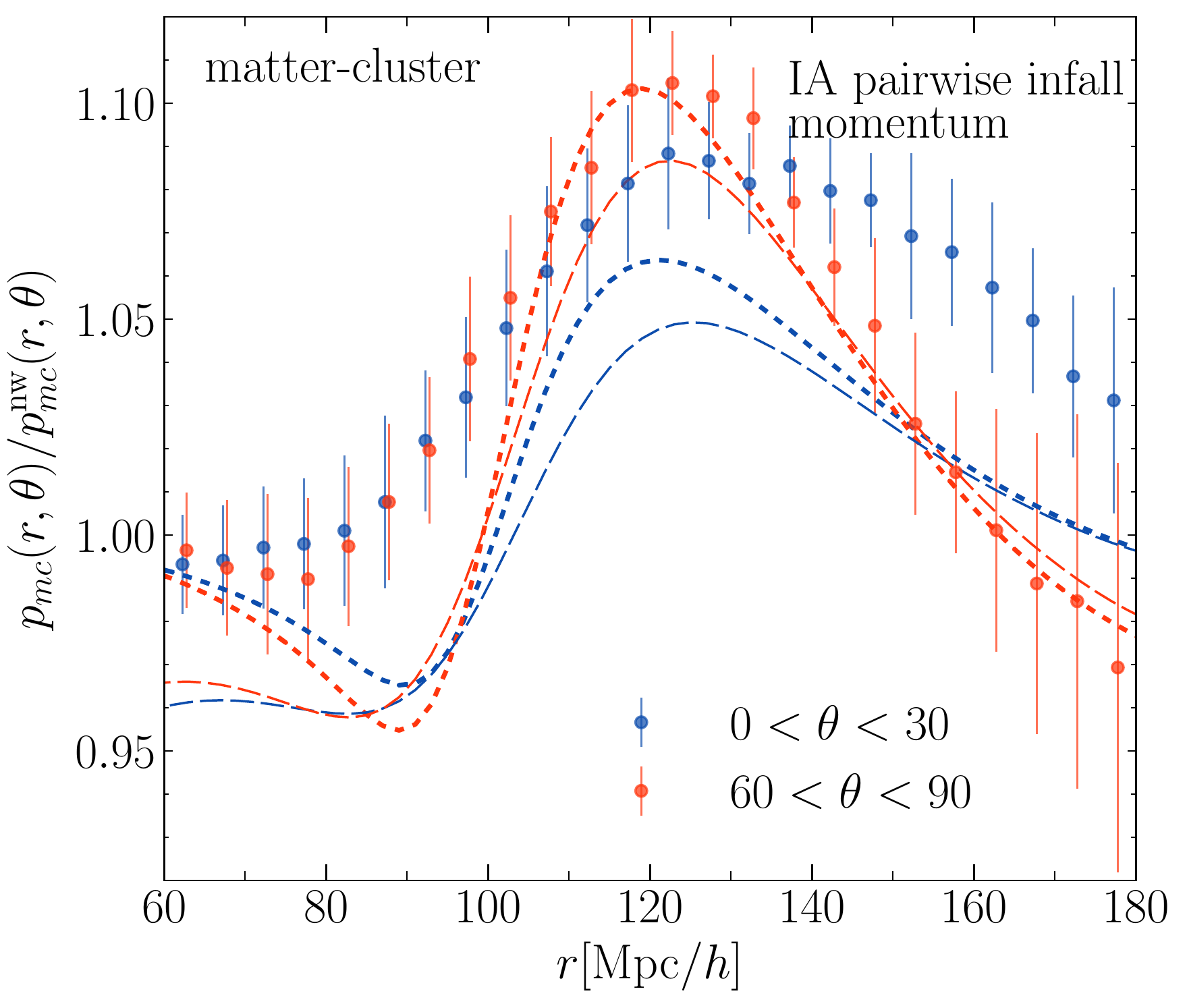}
\caption{
Ratio of the alignment mean infall momentum to that in linear theory without BAOs, $p_{mc}(r,\theta)/p_{mc}^{\rm nw}(r,\theta)$, where the denominator, $p_{mc}^{\rm nw}$, is the linear power spectrum without BAOs computed using the fitting formula \cite{Eisenstein:1998}. The numerator, $p_{mc}(r,\theta)$, is computed from the simulations (points), the LA model (dotted curves) and the NLA model (dashed curves). The blue and red points/curves are the results for $0<\theta<30^\circ$ and $60<\theta<90^\circ$, respectively. The blue and red points have been offset along the horizontal direction, $\pm 0.25\himpc$, respectively, for clarity.
}
\label{fig:vel_ia_bao_m_nw_z014}
\end{figure}

Finally, we move on to the alignment velocity statistics. Because there is no linear-level contribution to the IA of the momentum correlation function and pairwise velocity dispersion, we consider only the alignment pairwise infall momentum. Fig.~\ref{fig:vel_ia_bao_m_nw_z014} shows the ratios of the alignment mean infall momentum to that without BAOs, $p_{mc}(r,\theta)/p_{mc}^{\rm nw}(r,\theta)$. The numerator is computed from the simulations (points), the LA model (dotted curves) and the NLA model (dashed curves), while the denominator is from the LA model prediction with BAOs smeared out using the fitting formula of Ref.~\cite{Eisenstein:1998}.

Since the IA mean infall momentum consists of the conventional mean infall momentum and the VI correlation function, the small offsets seen in Fig.~\ref{fig:vel_nia_bao_m_z014} would affect the result. Nevertheless, the enhancement of BAO features for the separation perpendicular to the major axes of clusters is qualitatively captured by the LA and NLA models as shown by the red curves. On the other hand, BAO features for the IA mean infall momentum are suppressed along the major axis of a cluster due to the negative contribution of the VI correlation to BAOs. Thus, the BAO signals for $\theta\sim 0$ are less prominent, though the deviation of the measurement from the predictions is at most $\sim 3\%$. 

A more detailed modeling of the IA velocity statistics and their cosmological impacts will be studied in our future papers.

\section{Discussion and conclusions}\label{sec:conclusion}
In the literature the IA of galaxy/halo orientations have been studied in detail only with regard to the surrounding overdensity field. In this paper we focused on the statistics of IA characterized in phase space with density and velocity fields rather than the traditional one in three-dimensional position space with just the density field. For this purpose, we considered various velocity statistics, including the density-weighted VI correlation function, alignment pairwise infall momentum, alignment momentum correlation function, and density-weighted alignment pairwise velocity dispersion.

We derived simple analytic formulas for these velocity statistics under the assumption that the density fluctuation is a random Gaussian field and the velocity and tidal fields are related to the density by linear theory. The alignment mean infall momentum, $p_{AB}(r,\theta)$ [Eq.~(\ref{eq:infall_momentum})], momentum correlation function $\psi_{AB}(r,\theta)$ [Eq.~(\ref{eq:momentum_momentum})], and density-weighted pairwise velocity dispersion $\Sigma_{AB}^2(r,\theta)$ [Eq.~(\ref{eq:velocity_disp})] are expressed in terms of the GI and VI correlation functions with the $\cos{(n\theta)}$ terms with $n$ even as well as the conventional density and velocity correlation functions. The GI and VI functions in the formulas have been computed based on the LA model. We tested our theoretical models of velocity statistics by comparing them to the measurements of large-volume $N$-body simulations. We then found that our formula can explain the large-scale alignment signals in the measured mean infall momentum beyond $100\himpc$ up to $\sim 600\himpc$.

We have also studied how signatures of BAOs are imprinted into the velocity statistics and the IA in them. We detected the BAO signals in the conventional mean infall momentum, momentum correlation function, and pairwise velocity dispersion, which confirms the earlier measurement in the density-momentum and momentum-momentum power spectra in Fourier space in Ref.~\cite{Okumura:2014}.

In the VI correlation function, the contributions of BAOs are negative and they appear as a trough rather than a bump as is the case in the GI correlation function. Thus, the alignment pairwise infall momentum perpendicular and parallel to the major axes of clusters has a more and less prominent BAO bump, respectively, than the conventional pairwise infall momentum. This feature has been predicted by our models to some extent. 

The alignment pairwise momentum, $p_{AB}(r,\theta)$, or its ratio with the density correlation, may serve as a powerful tool to probe inflation. Future kSZ surveys will enable us to measure the large-scale velocity field and constrain cosmological models using the higher-order velocity moments \cite{Sugiyama:2017}. A possible contaminant of utilizing kSZ surveys is the effect of the optical depth. However, a promising method of measuring it has recently been proposed based on a semianalytic technique calibrated with x-ray observations \cite{Flender:2017}. 

In this paper we have presented the measurements of IA statistics and their theoretical modelings only in real space. Note that, however, in observations such as peculiar velocity and kSZ surveys, the velocity field is also sampled in redshift space, and thus is affected by redshift-space distortions (RSDs) \cite{Kaiser:1987}, as formulated in Refs.
~\cite{Okumura:2014,Sugiyama:2016,Howlett:2019}. The effect of RSDs on the alignment clustering statistics has been studied in Refs.~\cite{Hirata:2009,Martens:2018} in a different context. A preliminary analysis of RSDs in IA statistics has been performed in Ref.~\cite{Okumura:2017a}. It is straightforward to extend the presented modeling of velocity IA by the LA model in real space to that in redshift space. On the other hand, it is important to model the nonlinearity of IA statistics in order to maximize the encoded cosmological information \cite{Blazek:2015}. A detailed modeling of the effects of nonlinear RSDs on the IA statistics as well as the higher-order multipoles will be presented in our future work 

On very large scales, the amplitude of clustering of the density field is known to be affected by the effect of weak gravitational lensing \cite{Matsubara:2000}. However, such an effect on the velocity statistics has not been considered yet, while we are interested in the mean infall momentum on extraordinarily large scales. We will study of the impact of lensing effect on velocities including intrinsic alignments in our future work.

\begin{acknowledgements} We thank Keiichi Umetsu and Ken Osato for their contributions to the companion papers \cite{Okumura:2017a,Okumura:2018a}. T.~O. thanks Aniket Agrawal for useful conversations. T.~O. acknowledges support from the Ministry of Science and Technology of Taiwan under Grants No. MOST 106-2119-M-001-031-MY3 and the Career Development Award, Academia Sinina (AS-CDA-108-M02) for the period of 2019 to 2023. 
A.~T. was supported in part by MEXT/JSPS KAKENHI Grants No. JP15H05899 and No. JP16H03977.
T.~N. was supported by Japan Science and Technology Agency CREST JPMHCR1414, and by JSPS KAKENHI Grant Number JP17K14273. 
This research was supported by the World Premier International Research Center Initiative (WPI), MEXT, Japan. 
Numerical computations were carried out on Cray XC30 and XC50 at the Center for Computational Astrophysics, National Astronomical Observatory of Japan.
\end{acknowledgements}

\appendix
\section{Integral formulas for conditional average of Gaussian random fields}\label{sec:formula}

In this appendix, we present the formulas for the conditional average of the Gaussian fields, which are used to derive analytical expressions for the IA statistics of density and velocity fields in Sec.~\ref{sec:prediction}. 

Our goal is to derive useful analytical formulas to compute the conditional average given in Eq.~(\ref{eq:conditional_average_integral}). To do so, as a first step, we define another conditional average, fixing both of the variables $\gamma_0$ and $\theta$. Denoting it by $\langle F | \gamma_0,\theta \rangle$, we have 
\bey
\langle F | \gamma_0,\,\theta \rangle &\equiv& \frac{1}{(2\pi)^{3}|\det{\bf C}|^{1/2}} \nn \\ 
&\times &\int \prod^4_{a=1}d q_a F \exp{\left(-\frac{1}{2}q_i C^{-1}_{ij} q_j\right)}, \label{eq:gaussian_new}
\eey
where $\vq$ is given by Eq. (\ref{eq:vector_q}) and $\vC$ is its covariance matrix. In what follows, unless explicitly mentioned, subscripts $i,j$ will run over the range $1-6$, while $a,b$ run from $1$ to $4$, i.e., $1\leq i,j\leq 6$ and $1\leq a,b \leq 4$.  Given Eq.~(\ref{eq:gaussian_new}), the conditional average for the alignment statistics, given in Eq.~(\ref{eq:conditional_average_integral}), is computed by further integrating it over $\gamma_0$: 
\bey
\langle F | \theta\rangle = 2\pi \int_0^\infty  \langle F| \gamma_0,\,\theta \rangle \gamma_0 d\gamma_0. \label{eq:gaussian_new_integral}
\eey

In this appendix, we derive several analytical formulas for Eq. (\ref{eq:gaussian_new}). Below, for notational simplicity, we adopt the Einstein summation convention, and an index variable, such as $q_i$ and $q_a$, that appears twice in a single term implies summation over the index. Let us first decompose the exponent in Eq.~(\ref{eq:gaussian_new}) as
\bey
-\frac{1}{2}q_i C^{-1}_{ij} q_j &=& -\frac{1}{2}q_a Q_{ab} q_b \nn \\
&&+ A_a(q_5',q_6')\, q_a + B(q_5',q_6'),
\eey
Here, $q_5'$ and $q_6'$ are the new variables transformed from $q_5$ and $q_6$.  To be explicit, $(q_5',q_6')=(\gamma_0,\theta)$ (see Sec.~\ref{sec:gaussian_random}). The matrix $Q_{ab}$ is the submatrix of $C_{ij}^{-1}$, namely, $Q_{ab}=C_{ab}^{-1}$, and $A_a$ and $B$ are given by 
\bey
A_a &=& -C_{a5}^{-1}q_5 - C_{a6}^{-1}q_6 \nn \\
&=& -\gamma_0 \left( C_{a5}^{-1}\cos {2\theta}+C_{a6}^{-1}\sin {2\theta} \right), \\
B &=& -\frac{1}{2}\sum_{i,j=5}^6 C_{ij}^{-1}q_iq_j \nn \\
&=& \frac{\gamma_0^2}{2}\left( C_{55}^{-1}\cos^2{2\theta} + C_{66}^{-1}\sin^2{2\theta} +C_{56}^{-1}\sin{4\theta}\right). \ \ \ \ \ \ 
\eey
In order to analytically perform the multivariate Gaussian integral, we next change the integration variables from $q_a$, by introducing new variables $y_a$:
\be
y_a = Q_{ab}q_b - A_a. 
\ee

Using $\prod_{a=1}^4 dq_a = |\det{\bf Q}|^{-1} \prod_{a=1}^4 dy_a$, Eq. (\ref{eq:gaussian_new}) is recast as
\bey
\langle F | \gamma_0,\,\theta \rangle &=& \frac{1}{(2\pi)^{3}|\det{\bf C}|^{1/2}|\det{\bf Q}|} 
\exp{\left(\frac{1}{2}A_a Q^{-1}_{ab} A_b + B\right)} \nn \\
& &\times \int \prod^4_{c=1}d y_c \,\,F \,\,\exp{\left(-\frac{1}{2}y_a Q^{-1}_{ab} y_b\right)}. \label{eq:gaussian_new2}
\eey
Then, for the quantity $F$ given as the polynomial forms of $q_c$, the Gaussian integral in Eq.~(\ref{eq:gaussian_new2}) can be analytically performed. 

The simplest example is $F=q_a$. The final expression becomes
\bey
\langle q_a | \gamma_0,\,\theta \rangle &=& \frac{Q_{ab}^{-1}A_b}{2\pi |\det{\bf C}|^{1/2}|\det{\bf Q}|^{1/2}}  \nn \\
&& \times \exp{\left(\frac{1}{2}A_a Q^{-1}_{ab} A_b + B\right)} . \label{eq:formula1}
\eey
The second simplest case is to set $F=q_aq_b$. We then obtain an analytical expression of the form
\begin{widetext}
\be
\langle q_aq_b | \gamma_0,\theta \rangle = \frac{1}{2\pi |\det{\bf C}|^{1/2}|\det{\bf Q}|^{1/2}}  
Q_{ac}^{-1}Q_{bd}^{-1}\left(Q_{cd}+A_cA_d\right) 
\exp{\left(\frac{1}{2}A_a Q^{-1}_{ab} A_b + B\right)}. \label{eq:formula2}
\ee
Likewise, the cases for higher-order polynomials can also be computed, and we have
\bey
\langle q_aq_bq_c | \gamma_0,\theta \rangle= \frac{1}{2\pi |\det{\bf C}|^{1/2}|\det{\bf Q}|^{1/2}}  
Q_{ad}^{-1}Q_{be}^{-1}Q_{cf}^{-1}
\left(Q_{de}A_f +Q_{df}A_e+Q_{ef}A_d +A_dA_eA_f\right) 
\exp{\left(\frac{1}{2}A_a Q^{-1}_{ab} A_b + B\right)}, \ \ \ \  \label{eq:formula3}
\eey 
and 
\bey
\langle q_aq_bq_cq_d | \gamma_0,\,\theta \rangle&=& \frac{1}{2\pi |\det{\bf C}|^{1/2}|\det{\bf Q}|^{1/2}}  
\left( Q_{ab}^{-1}Q_{cd}^{-1} +Q_{ac}^{-1}Q_{bd}^{-1} +Q_{ad}^{-1}Q_{bc}^{-1} 
+  Q_{ab}^{-1} U_cU_d +Q_{ac}^{-1} U_bU_d +Q_{bc}^{-1} U_aU_d \right. \ \ \ \  \nn \\
&& \left.+  Q_{ad}^{-1} U_bU_c +Q_{bd}^{-1} U_aU_c +Q_{cd}^{-1} U_aU_b 
+   U_aU_bU_cU_d \right)
\exp{\left(\frac{1}{2}A_a Q^{-1}_{ab} A_b + B\right)}, \label{eq:formula4}
\eey
\end{widetext}
where we defined $U_a\equiv Q_{ab}^{-1}A_b$. 

The formulas given above are used to further compute the conditional average of Eq.~(\ref{eq:gaussian_new_integral}), together with the covariance matrix given in Eq.~(\ref{eq:covariance}). 
The remaining calculations we need to perform are just one-dimensional Gaussian integrals.
Thus, the final integration can also be performed analytically. Hence, by applying Eqs.~(\ref{eq:formula1})--(\ref{eq:formula4}) to each term of the alignment statistics of density and velocity fields [Eqs.~(\ref{eq:acf_density}) and (\ref{eq:infall_momentum}) -- (\ref{eq:velocity_disp})], the analytical expressions summarized in Sec.~\ref{sec:prediction} can be obtained through a straightforward algebraic manipulation, which can be done easily with mathematical software such as \textit{Mathematica}\footnote{https://www.wolfram.com/mathematica/}.


 \end{document}